# Geometric Phase Effect in Thermodynamic Properties and in the Imaginary-Time Multi-Electronic-State Path Integral Formulation

*Yu Zhai* †, *Youhao Shang* †, *Jian Liu* *

Beijing National Laboratory for Molecular Sciences, Institute of Theoretical and Computational Chemistry, College of Chemistry and Molecular Engineering,

Peking University, Beijing 100871, China

AUTHOR INFORMATION

Corresponding Author

* Electronic mail: jianliupku@pku.edu.cn

Author Contributions

† Y. Z. and Y. S. contributed equally.

ABSTRACT

The geometric phase (GP) is a fundamental quantum effect arising from conical intersections (CIs), with profound consequences for vibronic energy levels. Standard imaginary-time path integral molecular dynamics (PIMD) based on the Born-Oppenheimer approximation does not account for the GP, potentially leading to significant errors in low-temperature thermodynamic properties. In this Perspective, we demonstrate that the multi-electronic-state path integral (MES-PI) formulation in imaginary time—developed in *J. Chem. Phys.* **2018**, 148, 102319—naturally captures the GP effect through the electronic trace of the product of statistically weighted overlap matrices between successive imaginary-time slices. This crucial capability was already implicit in the benchmark MES-PIMD simulations in that foundational work. To isolate this topological effect from other nonadiabatic effects, we introduce a geometric signature matrix (for the CI) and a winding-number-induced phase factor, constructing an *ad hoc* GP-excluded MES-PI method. Comparing this *ad hoc* baseline against the rigorous MES-PI approach allows us to unambiguously quantify the impact of the GP on thermodynamic properties. While simpler approximations exist when only the ground electronic-state is considered, MES-PIMD is the most general and accurate approach applicable to real complex systems where the location and topology of CI seams are often not known *a priori.*

TOC GRAPHICS

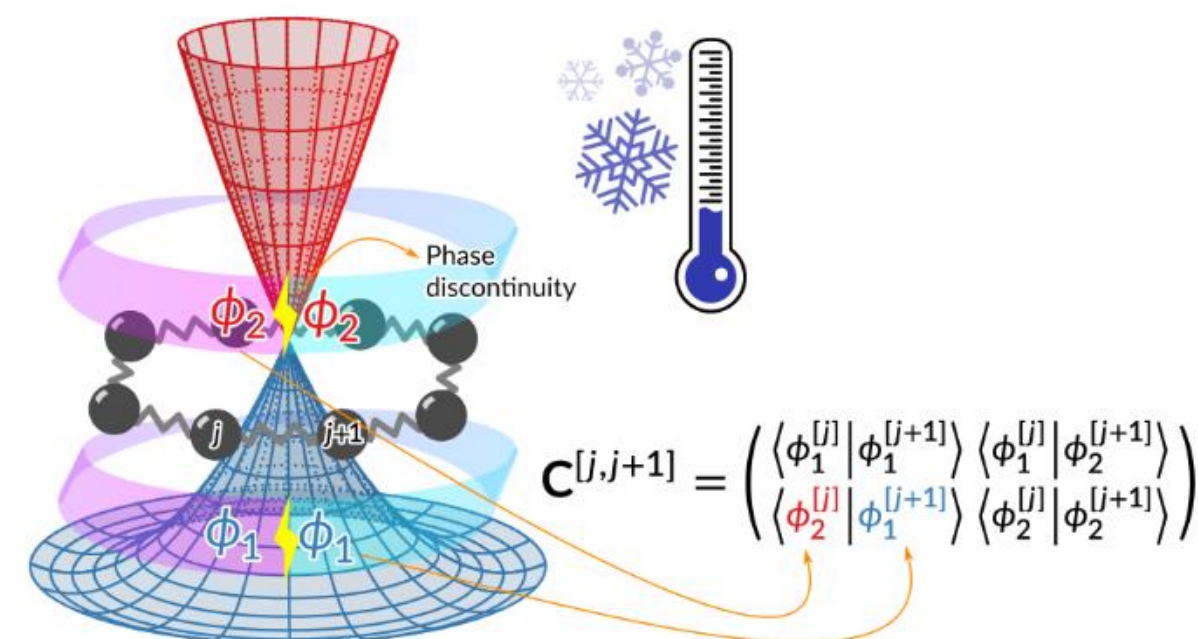

KEYWORDS

Multi-electronic-state path integral; geometric phase; conical intersection; path integral molecular dynamics; Single-electronic-state limit; geometric signature matrix

The geometric phase (GP) [1-4] in molecules, a special instance of the Berry phase [5, 6], manifests as a topological phase factor imposed on the nuclear wavefunction when the system traverses a closed loop encircling a conical intersection (CI) seam [7, 8]. This phenomenon arises specifically at the intersecting potential energy surfaces (PESs) of different adiabatic electronic states, where the breakdown of the Born-Oppenheimer (BO) approximation is most profound. The GP originates from the nontrivial topology of the adiabatic electronic basis: for real-valued Hamiltonians, the real-valued electronic wavefunction undergoes a sign change upon completing the loop, necessitating a compensating phase in the nuclear sector to preserve the single-valuedness of the total molecular wavefunction. In the limit of a complete electronic Hilbert space, this phenomenon is formally isomorphic to the Aharonov-Bohm effect, where a charged particle accumulates a quantum phase determined by the vector potential while encircling a magnetic flux line, occurring even along paths where the magnetic field vanishes. Consequently, the GP in molecules is also often referred to as the molecular Aharonov-Bohm effect [9].

The presence of GP profoundly influences nuclear dynamics, altering vibronic levels and interference patterns of reaction pathways encircling the CI. Experimental observations, in comparison to simulations for demonstrating the GP effect, include the paramagnetic resonance spectrum of copper acetate [10], the infrared spectrum of $Na_3$ [11, 12], and the differential scattering cross section of the $H + H_2 \rightarrow H_2 + H$ or $H + HD \rightarrow H_2 + D$ reaction [13-16]. The GP effect manifests in photodissociation reactions[17-19]. In addition, simulations have shown that the GP effect near conical intersections suppresses the transfer rate[20] (related to electronic frustration[20-22]) or reaction rate[23]. Most of these previous demonstrations of the GP effect focused on dynamic and spectroscopic properties. In this Perspective, we first explicitly show that the GP also has a significant impact on low-temperature thermodynamic properties.

To rigorously quantify the GP effect on thermodynamics, we utilize the electronically adiabatic representation and the imaginary-time multi-electronic-state path integral (MES-PI) formulation [24, 25]. When MES-PI is performed by molecular dynamics (MES-PIMD) with reweighting approaches, where the reweighting factors may be negative as described in ref [24], it offers a practical and exact approach for simulating nonadiabatic molecular systems with a large number of nuclear degrees of freedom (DOFs). Crucially, while the MES-PIMD approach is representation-independent, the explicit manifestation of GP—and the challenges in capturing it—appear naturally in the adiabatic representation of the prevailing Born-Huang interpretation [26] of real molecular systems. In this Perspective, we demonstrate the GP effect in thermodynamic properties and investigate how MES-PIMD naturally captures the GP *via* the electronic trace of the product of statistically-weighted overlap matrices along the ring polymer.

**Jahn-Teller $E\otimes e$ Model**. We employ the archetypal "Mexican hat" Jahn-Teller $E\otimes e$ model [27-30] to investigate the GP effect. The model involves two electronic states of $E$ symmetry and two nuclear DOFs of $e$ symmetry. It commonly serves as a prototypical system for studying the Jahn-Teller effect[31-33] in a wide range of molecular systems, such as $H_3$ [14, 34], $H_3^+$ [35], $NH_3^+$ [36], Tutton's salt[37], the $Cu_5$ cluster[38], and crystal systems, such as $Ba_2MgReO_6$ [39]. In the model, both the mass and reduced Planck constant $\hbar$ are set to 1. The Hamiltonian in the diabatic representation reads

$$\hat{\mathbf{H}}_{\text{diab}} = \hat{\mathbf{K}} + \hat{\mathbf{V}}, \tag{1}$$

where the nuclear kinetic energy is

$$\hat{\mathbf{K}} = \frac{1}{2}\hat{\mathbf{P}}\cdot\hat{\mathbf{P}}, \tag{2}$$

with the nuclear momentum operator $\hat{\mathbf{P}} = -\mathrm{i}\nabla$, and the derivatives apply on not only the nuclear

DOFs, but also the electronic basis set [40]. Because the diabatic electronic basis set is independent of nuclear coordinates $\mathbf{R}$, the nuclear kinetic energy is a diagonal matrix in eq (2). The diabatic potential energy matrix (DPEM) of eq (1) is

$$\mathbf{V} = \frac{1}{2}\left(x^2 + y^2 + c^2\right)\mathbf{1}_{\text{ele}} + c\begin{pmatrix} -x & y \\ y & x \end{pmatrix} = \frac{1}{2}\left(r^2 + c^2\right)\mathbf{1}_{\text{ele}} + cr\begin{pmatrix} -\cos\theta & \sin\theta \\ \sin\theta & \cos\theta \end{pmatrix}. \tag{3}$$

Hereafter, for convenience, we employ polar coordinates $(r,\theta)$, with $\theta \in [0,2\pi)$. In eq (3), the harmonic-oscillator frequency of the Jahn-Teller $E \otimes e$ model [27-30] is set to $\omega = 1$. We also set $\hbar = M = 1$ for the reduced Planck constant and the mass. Setting these three parameters to unity establishes the reduced units (r.u.) employed throughout this Perspective, while $c$ remains an adjustable parameter. The findings obtained in these reduced units can be readily mapped to real molecular systems. Diagonalization of DPEM yields the adiabatic representation. However, under the convention that the eigenvectors (i.e., adiabatic electronic wavefunctions for molecules) are real and single-valued, obtaining the nonadiabatic coupling matrix *via* brute-force differentiation precludes a correct description of the GP effect. This is because the topological phase information is hidden in the boundary conditions of the wavefunction rather than its local derivatives. Brute-force differentiation yields the (incorrect) Hamiltonian in the adiabatic representation

$$\hat{\mathbf{H}}_{\text{adia}}^{\text{BF}} = \hat{\mathbf{K}}_{\text{BF}} + \hat{\mathbf{\Lambda}}. \tag{4}$$

where the adiabatic potential energy surface (PES) matrix (depicted in Figure 1(a)) reads

$$\mathbf{\Lambda} = \begin{pmatrix} \frac{1}{2}\left(c-r\right)^2 & 0 \\ 0 & \frac{1}{2}\left(c+r\right)^2 \end{pmatrix}, \tag{5}$$

and the corresponding real, single-valued (electronic) eigenvectors are

$$\begin{pmatrix} \cos\theta/2 \\ -\sin\theta/2 \end{pmatrix}, \quad \text{and} \quad \begin{pmatrix} \sin\theta/2 \\ \cos\theta/2 \end{pmatrix}, \quad \text{with } \theta \in [0,2\pi). \tag{6}$$

The kinetic energy is then expanded as the summation of the canonical nuclear kinetic energy and the nonadiabatic coupling matrix

$$\hat{\mathbf{K}}_{\mathrm{BF}} = \frac{1}{2}\hat{\mathbf{P}}_{\mathrm{can}} \cdot \hat{\mathbf{P}}_{\mathrm{can}} + \begin{pmatrix} +\dfrac{1}{8r^2} & -\dfrac{1}{2r^2}\dfrac{\partial}{\partial\theta} \\ \dfrac{1}{2r^2}\dfrac{\partial}{\partial\theta} & +\dfrac{1}{8r^2} \end{pmatrix} , \tag{7}$$

Note that eq (7) has already been expressed in the adiabatic electronic basis set, where the canonical nuclear momentum operator $\hat{\mathbf{P}}_{\mathrm{can}} = -\mathrm{i}\nabla_{\mathrm{can}}$, the canonical nuclear kinetic energy operator $\frac{1}{2}\hat{\mathbf{P}}_{\mathrm{can}} \cdot \hat{\mathbf{P}}_{\mathrm{can}}$, and the nonadiabatic coupling matrix should *not* explicitly operate on the adiabatic electronic basis set [40]. In polar coordinates, the canonical nuclear kinetic energy operator reads

$$\frac{1}{2}\hat{\mathbf{P}}_{\mathrm{can}} \cdot \hat{\mathbf{P}}_{\mathrm{can}} = -\frac{1}{2}\left( \frac{\partial^2}{\partial r^2} + \frac{1}{r}\frac{\partial}{\partial r} + \frac{1}{r^2}\frac{\partial^2}{\partial\theta^2} \right)\hat{\mathbf{1}}_{\mathrm{ele}} \quad . \tag{8}$$

In the nuclear kinetic energy operator $\hat{\mathbf{K}}_{\mathrm{BF}}$, the diagonal element in the nonadiabatic coupling matrix $1/8r^2$ is the diagonal Born-Oppenheimer correction (DBOC) term. We note that under the $2\pi$-periodic boundary condition in $\theta$, the operation of $\partial / \partial\theta$ on the basis states yields purely imaginary values; consequently, $\hat{\mathbf{K}}_{\mathrm{BF}}$ is Hermitian.

Note that the real-valued and single-valued (electronic) eigenvectors in eq (6) acquire a geometric phase of $\pi$ (i.e., a phase factor of $-1$) upon traversing a closed path around the CI ($\theta \to \theta + 2\pi$) in the nuclear coordinate space, and are necessarily discontinuous at the branch cut (e.g., at $\theta = 0$) to remain single-valued. However, eq (7), the kinetic energy operator of eq (4), involves both first-order and second-order derivative coupling terms, which inherently request continuity of the (electronic) eigenvectors or electronic wavefunctions in the nuclear coordinate space. This is contradictory to the discontinuity of eq (6).

One way to resolve this contradiction is to apply a phase factor $\mathrm{e}^{\mathrm{i}k\theta/2}$, where $k$ is an arbitrary odd integer, to the eigenvectors in eq (6), such that they become single-valued and continuous [2, 18, 41]. The physical results are invariant with respect to the choice of odd integer $k$. Choosing the phase-factor-modified eigenvectors with $k=1$ as the adiabatic electronic basis set, without loss of generality, leads to the GP-included Hamiltonian in the adiabatic representation of the Hamiltonian of eq (1),

$$\hat{\mathbf{H}}_{\mathrm{adia}}^{\mathrm{GP}}=\hat{\mathbf{K}}_{\mathrm{GP}}+\hat{\mathbf{\Lambda}}, \tag{9}$$

where the nuclear kinetic energy is

$$\hat{\mathbf{K}}_{\mathrm{GP}}=\frac{1}{2}\hat{\mathbf{P}}_{\mathrm{can}}\cdot\hat{\mathbf{P}}_{\mathrm{can}}+\begin{pmatrix}\dfrac{1}{4r^2}-\dfrac{\mathrm{i}}{2r^2}\dfrac{\partial}{\partial\theta} & -\dfrac{\mathrm{i}}{4r^2}-\dfrac{1}{2r^2}\dfrac{\partial}{\partial\theta}\\ \dfrac{\mathrm{i}}{4r^2}+\dfrac{1}{2r^2}\dfrac{\partial}{\partial\theta} & \dfrac{1}{4r^2}-\dfrac{\mathrm{i}}{2r^2}\dfrac{\partial}{\partial\theta}\end{pmatrix} \tag{10}$$

with eq (8) for polar coordinates. The GP-included Hamiltonian (eq (9)) is equivalent to the diabatic one (eq (1)), while the GP-excluded Hamiltonian (eq (4)) is artificial. Analogous to eq (7), $\hat{\mathbf{K}}_{\mathrm{GP}}$ is Hermitian because the derivative operator $\partial/\partial\theta$ is anti-Hermitian. The Hamiltonians of eq (1) and eq (9) are related to each other by unitary transform $\mathbf{H}_{\mathrm{adia}}^{\mathrm{GP}}=\mathbf{T}^{\dagger}(\mathbf{R})\mathbf{H}_{\mathrm{diab}}\mathbf{T}(\mathbf{R})$, where the columns of $\mathbf{T}(\mathbf{R})$ are the phase-factor-modified eigenvectors with $k=1$.

We first establish exact benchmarks by solving the time-independent Schrödinger equation for both the GP-included Hamiltonian (eq (9)) and GP-excluded Hamiltonian (eq (4)) using a basis-set-expansion method. The cases with parameter $c=1\,\mathrm{r.u.}$ are illustrated in Figure 1 b-d. The value of parameter $c$ is chosen through an order-of-magnitude calibration against the characteristic physical quantities of the Jahn-Teller system $Ba_2MgReO_6$ reported in ref [39]. That is, $E_{\mathrm{JT}}=M\omega^2c^2/2\approx 80\,\mathrm{meV}$ and $\hbar\omega\approx 50\,\mathrm{meV}$ (used in ref [39]) yields the parameter value $c\,/\,\mathrm{r.u.} = \sqrt{2E_{\mathrm{JT}}\,/\,\hbar\omega} \approx 2$. In the vicinity of the conical-intersection point, namely, the origin ($r = 0$), the nuclear kinetic energy operator (either eq (7) or eq (10)) dominates the corresponding Hamiltonian, since its radial-derivative and $1/r^2$ terms diverge as $r \to 0$, while the potential energy (eq (5)) remains finite. Therefore, as described in Supporting Information (SI) Section S1.2, we expand the vibronic eigenstates using the basis set composed of the eigenfunctions of the nuclear kinetic energy operator in eq (10) [eq (7)], which enables efficient convergence of the eigenspectrum for the Hamiltonian in eq (9) [eq (4)]. These eigenfunctions of the nuclear kinetic-energy operator correctly capture the singular short-range behavior of the ground vibronic state in the GP-excluded case—most notably the cusp-like structure and the strictly vanishing probability density at the origin—as illustrated in Figure S3. The ground vibronic wavefunction is *not* differentiable at the origin (the CI point). In contrast, in the GP-included case, the doubly degenerate ground vibronic states exhibit smooth probability density at the origin, as shown in Figure S2. The degenerate ground vibronic wavefunctions are differentiable at the origin.

The difference between eigenenergy state structures can, in principle, be manifested in thermodynamic properties (Figure 1(c-d)). After we obtain the eigensystem (i.e., energy eigenvalues and eigenstates) as depicted in Figure 1(b), we evaluate thermodynamic properties by summation over states, i.e., for an arbitrary physical observable $\hat{B}$, $\langle B \rangle = \sum_{\nu} B_{\nu} \mathrm{e}^{-\beta E_{\nu}} \Big/ \sum_{\nu} \mathrm{e}^{-\beta E_{\nu}}$, where $\beta = 1/k_{\mathrm{B}}T$ is the inverse temperature, $k_{\mathrm{B}}$ is the Boltzmann constant, and $B_{\nu} = \langle \Psi_{\nu} | \hat{B} | \Psi_{\nu} \rangle$ is the expectation value of the $\nu$-th vibronic eigenstate. Although the overall trends of the eigenenergy spectra are similar, as shown in Figure 1(b), the most significant difference is that the ground state is doubly degenerate in the GP-included case, whereas it is non-degenerate in the GP-excluded case, as already well-known in the literature on the Jahn-Teller effects [12, 42, 43]. It manifests in the heat capacity $C_V$, especially at low temperatures (large $\beta$), where the GP-included case exhibits distinct behavior compared to the GP-excluded case (Figure 1(d)).

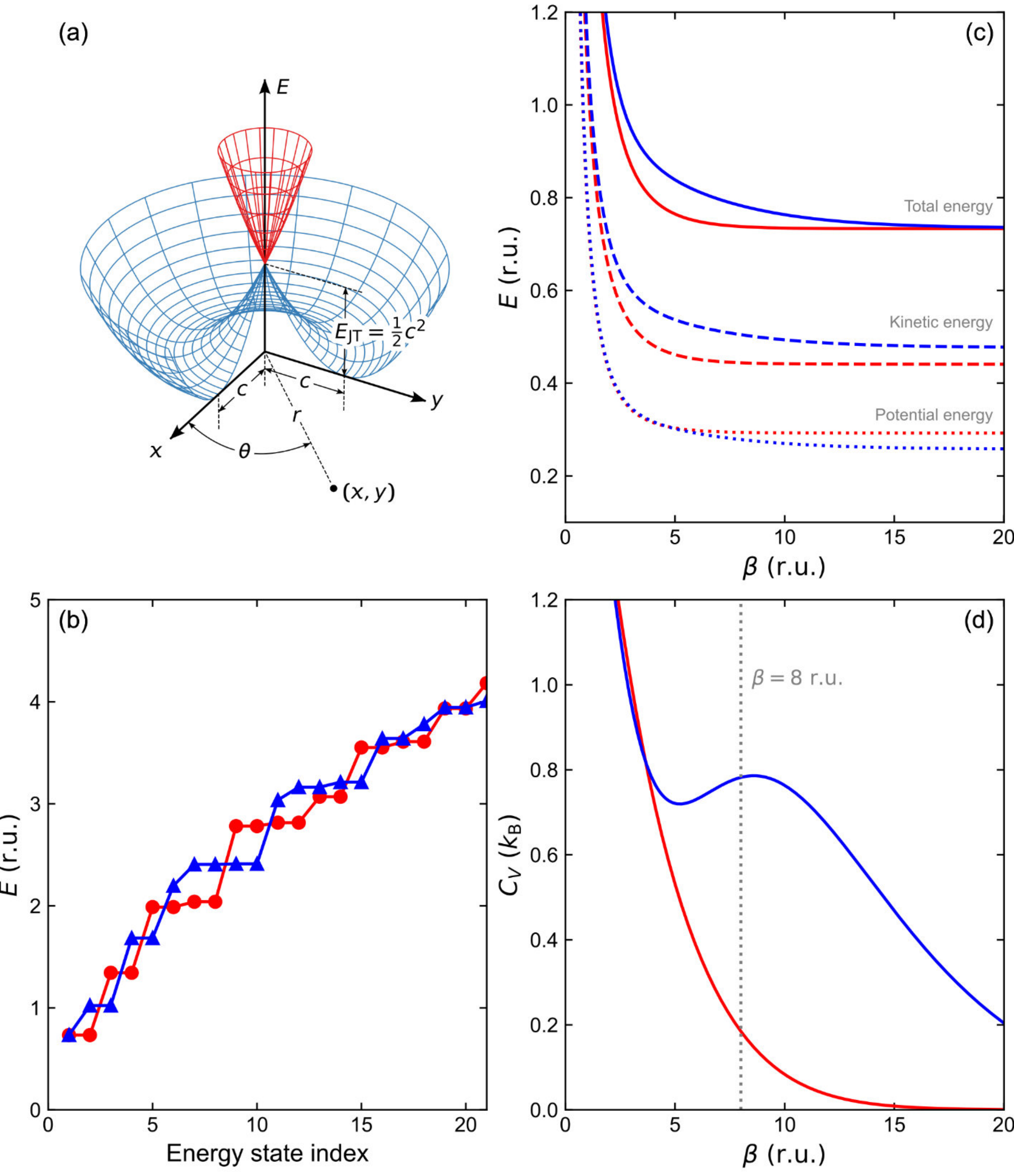

**Figure 1.** (a) Adiabatic potential energy surfaces of the Jahn-Teller $E \otimes e$ model (eq (1)). (b-d) Results obtained by directly solving the time-independent Schrödinger equation by the basis-set-expansion method. (b) Eigenenergy states; (c), (d) Physical properties at different temperatures. In (b-d), $c = 1$ r.u.. Red lines: GP-included case; Blue lines: GP-excluded case.

**Multi-Electronic-State Path Integral Formulation**. We now consider the imaginary time path integral formulation, which is exact as well as practical for evaluating thermodynamic properties of large systems. The partition function of a general multi-electronic-state (MES) system is $Z = \mathrm{Tr}_{\mathrm{n,e}}[\mathrm{e}^{-\beta\hat{\mathbf{H}}}]$, where $\mathrm{Tr}_{\mathrm{n,e}}$ represents the trace over both the nuclear and electronic DOFs. While the trace over the electronic DOFs is performed by the sum of all electronic states, the trace over the nuclear DOFs is accomplished by the integral of the nuclear coordinate space. When the diabatic representation is used, after $N_{\mathrm{bead}} - 1$ substitutions of the resolution of identity of the nuclear space, e.g., $\hat{\mathbf{1}}_{\mathrm{nuc}}^{[j]} = \int \mathrm{d}\mathbf{R}^{[j]} \,|\mathbf{R}^{[j]}\rangle\langle\mathbf{R}^{[j]}|$ with the nuclear coordinate vector, $\mathbf{R}^{[j]}$, at the *j*-th imaginary-time slice, i.e., the *j*-th bead, the Trotter decomposition form [44, 45] of $\mathrm{e}^{-\beta\hat{\mathbf{H}}}$ leads to the diabatic-MES-PI representation of the partition function[24, 25],

$$Z_{N_{\mathrm{bead}}}^{\mathrm{diab}} = \left|\frac{N_{\mathrm{bead}}\mathbf{M}}{2\pi\beta\hbar^2}\right|^{\frac{N_{\mathrm{bead}}}{2}} \int\cdots\int \mathrm{d}\{\mathbf{R}^{[j]}\}\mathrm{e}^{-\beta U_{\mathrm{spr}}}\mathrm{Tr}_{\mathrm{e}}\left[\prod_{j=1}^{N_{\mathrm{bead}}} \mathrm{e}^{-\frac{\beta}{N_{\mathrm{bead}}}\mathbf{V}^{[j]}}\right], \tag{11}$$

where $\mathbf{M}$ is the diagonal mass matrix, with each diagonal element being the mass of the corresponding dimension; the spring potential $U_{\mathrm{spr}} \equiv U_{\mathrm{spr}}\left(\left\{\mathbf{R}^{[j]}\right\}\right)$ is

$$U_{\mathrm{spr}}\left(\left\{\mathbf{R}^{[j]}\right\}\right) = \frac{N_{\mathrm{bead}}}{2\beta^2\hbar^2}\sum_{j=1}^{N_{\mathrm{bead}}}\left(\mathbf{R}^{[j]} - \mathbf{R}^{[j+1]}\right)^{\mathrm{T}}\mathbf{M}\left(\mathbf{R}^{[j]} - \mathbf{R}^{[j+1]}\right), \tag{12}$$

which comes from the nuclear kinetic energy operator; and $\mathbf{V}^{[j]} = \mathbf{V}\left(\mathbf{R}^{[j]}\right)$ is the DPEM of the $j$-th bead. In eqs (11) and (12), $\mathbf{R}^{[N_{\text{bead}}+1]} \equiv \mathbf{R}^{[1]}$. Equation (11) for MES systems involves a closed path in nuclear coordinate space, discretized into $N_{\text{bead}}$ beads and returning to its starting point after completing the loop, thereby forming a ring polymer. In eq (12), $\mathbf{R}^{[j]}$ and $\mathbf{R}^{[j+1]}$ are the nuclear coordinate vectors at two successive time slices, i. e., two successive beads along the ring polymer. $Z^{\text{diab}}_{N_{\text{bead}}}$ (eq (11)) converges to the exact partition function $Z$ in the limit $N_{\text{bead}} \to \infty$. A graphical representation of the diabatic-MES-PI framework (in eq (11)) is a closed "ring polymer" with beads linked by harmonic springs originating from the nuclear kinetic energy operator, as defined in eq (12), analogous to the conventional single-electronic-state case[46-49]. The crucial distinction is that the scalar potential energy is replaced by a matrix-valued diabatic potential energy operator (i.e., DPEM). Consequently, the statistical weight of the ring polymer is determined not by a sum of scalar energies but by the trace of the imaginary-time-ordered product of matrix exponentials of the diabatic potential energy along the path.

The usual method to evaluate $\exp(-\beta\mathbf{V}^{[j]}/N_{\text{bead}})$ of eq (11) is to use the equality, $\mathbf{V}^{[j]}\mathbf{T}^{[j]} = \mathbf{T}^{[j]}\mathbf{\Lambda}^{[j]}$, where $\mathbf{\Lambda}^{[j]} = \mathbf{\Lambda}\left(\mathbf{R}^{[j]}\right)$ is a diagonal matrix whose diagonal elements are the eigenvalues of $\mathbf{V}^{[j]}$, and $\mathbf{T}^{[j]} = \mathbf{T}\left(\mathbf{R}^{[j]}\right)$ is a unitary matrix whose column vectors are the eigenvectors of $\mathbf{V}^{[j]}$. This corresponds to the adiabatic representation of the same system

$$\mathrm{e}^{-\frac{\beta}{N_{\text{bead}}}\mathbf{V}\left(\mathbf{R}^{[j]}\right)} = \mathbf{T}^{[j]}\mathrm{e}^{-\frac{\beta}{N_{\text{bead}}}\mathbf{\Lambda}^{[j]}}\mathbf{T}^{[j]\dagger} \quad , \tag{13}$$

where $\mathbf{T}^{[j]\dagger}$ is the conjugate transpose of $\mathbf{T}^{[j]}$. The *m*-th column vector of matrix $\mathbf{T}^{[j]}$ represents the *m*-th adiabatic electronic state $\left|\phi_m^{[j]}\right\rangle = \left|\phi_m\left(\mathbf{R}^{[j]}\right)\right\rangle$. Substitution of eq (13) into eq (11) yields the *adiabatic-MES-PI* representation of the partition function[24]:

$$Z_{N_{\text{bead}}}^{\text{adia}} = \left|\frac{N_{\text{bead}}\mathbf{M}}{2\pi\beta\hbar^2}\right|^{\frac{N_{\text{bead}}}{2}} \int\cdots\int \mathrm{d}\left\{\mathbf{R}^{[j]}\right\} \mathrm{e}^{-\beta U_{\text{spr}}} \mathrm{Tr}_{\mathrm{e}}\left[\prod_{j=1}^{N_{\text{bead}}} \mathrm{e}^{-\frac{\beta}{N_{\text{bead}}}\boldsymbol{\Lambda}^{[j]}} \mathbf{C}^{[j,j+1]}\right], \tag{14}$$

where $\mathbf{C}^{[j,j+1]} \equiv \mathbf{C}\left(\mathbf{R}^{[j]},\mathbf{R}^{[j+1]}\right) = \mathbf{T}^{[j]\dagger}\mathbf{T}^{[j+1]}$ is the overlap matrix of adiabatic electronic states of two successive beads along the ring polymer. For instance, the entry at row $m$, column $n$ of overlap matrix $\mathbf{C}^{[j,j+1]}$ is $C_{mn}^{[j,j+1]} = \left\langle \phi_m^{[j]} \middle| \phi_n^{[j+1]} \right\rangle = \left\langle \phi_m\left(\mathbf{R}^{[j]}\right) \middle| \phi_n\left(\mathbf{R}^{[j+1]}\right) \right\rangle$. The term "overlap matrix $\mathbf{C}\left(\mathbf{R}^{[j]},\mathbf{R}^{[j+1]}\right) = \mathbf{T}^{[j]\dagger}\mathbf{T}^{[j+1]}$" (between two successive time slices) was first **explicitly** defined in the main text and Appendix A of ref [24]. Overlap matrices were intrinsically involved in both *adiabatic-MES-PI* and the diagonalization approach of *diabatic-MES-PI* in ref [24].

Equation (14) of the *adiabatic-MES-PI* representation requests the knowledge of the adiabatic electronic states of each bead. The non-adiabatic couplings are inherently included in the overlap matrix, $\mathbf{C}^{[j,j+1]}$. In addition to overlap matrices $\left\{\mathbf{C}^{[j,j+1]}\right\}$, the sampling of MES-PIMD requires only the adiabatic PESs (eq (5)), the spring potential (eq (12)), and their first-order derivatives with respect to nuclear coordinates. Remarkably, MES-PIMD does *not* require any knowledge of CIs, the continuous complex electronic basis set, or the correct formula of the nuclear kinetic energy operator (such as eq (10) for the present model), which is too computationally expensive to be feasible for most conventional electronic structure calculations for large molecular systems.

As previously stated, directly employing the Hamiltonian of eq (4) can lead to omission of the GP effect. Importantly, even when the adiabatic basis set is real-valued and globally single-valued but exhibits discontinuity (due to branch cuts), the adiabatic-MES-PI representation **does capture** the GP effect! We have shown that eqs (11) and (14) are equivalent [24], so the adiabatic-MES-PI representation of the partition function, $Z_{N_{\text{bead}}}^{\text{adia}}$, converges to the correct partition function in the limit $N_{\text{bead}} \to \infty$ , just as $Z_{N_{\text{bead}}}^{\text{diab}}$ does. The GP is naturally included in $\text{Tr}_{\text{e}}\left[\prod_{j=1}^{N_{\text{bead}}}\left(\exp(-\beta\mathbf{\Lambda}^{[j]}/N_{\text{bead}})\mathbf{C}^{[j,j+1]}\right)\right]$, the electronic trace of the product of statistically weighted overlap matrices along the ring polymer. $\mathbf{T}^{[j]}$ can be determined up to a set of phase factors for all its electronic states or eigenvectors. The trace over the electronic states in eq (14) is not altered with the phase factors, since $\mathbf{T}^{[j]}$ appears in both $\mathbf{C}^{[j-1,j]} = \mathbf{T}^{[j-1]\dagger}\mathbf{T}^{[j]}$ and $\mathbf{C}^{[j,j+1]} = \mathbf{T}^{[j]\dagger}\mathbf{T}^{[j+1]}$, the product $\mathbf{C}^{[j-1,j]}\exp(-\beta\mathbf{\Lambda}^{[j]}/N_{\text{bead}})\mathbf{C}^{[j,j+1]}$ remains the same when the phase of any column vector of $\mathbf{T}^{[j]}$ is varied. When *real-valued* electronic wavefunctions of adiabatic electronic states are used for evaluating overlap matrix $\mathbf{C}^{[j,j+1]}$, one may brute-forcely align the plus/minus sign of the column vector of matrix $\mathbf{T}^{[j]}$, or equivalently the phase of the electronic wavefunction of the $j$-th bead, and that of the corresponding column vector of matrix $\mathbf{T}^{[j+1]}$. However, when real-valued electronic wavefunctions are discontinuous around the CI, it is impossible to accomplish such an alignment in overlap matrix $\mathbf{C}^{[j,j+1]}$ for each pair of successive beads $[j, j+1]$ along the ring polymer. Therefore, in practice, we can directly use (discontinuous) real-valued (electronic) eigenvectors obtained from the brute-force diagonalization of a model Hamiltonian of the diabatic representation. This suggests that, even when overlap matrices $\left\{\mathbf{C}^{[j,j+1]}\right\}$ are constructed directly from discontinuous real-valued electronic wavefunctions provided by most conventional electronic structure software, the *adiabatic-MES-PI* representation fully captures the GP effect. Consequently, it yields correct results for thermodynamic properties, provided that the full overlap matrix is retained. Figure 2(a) depicts a schematic representation of adiabatic-MES-PI.

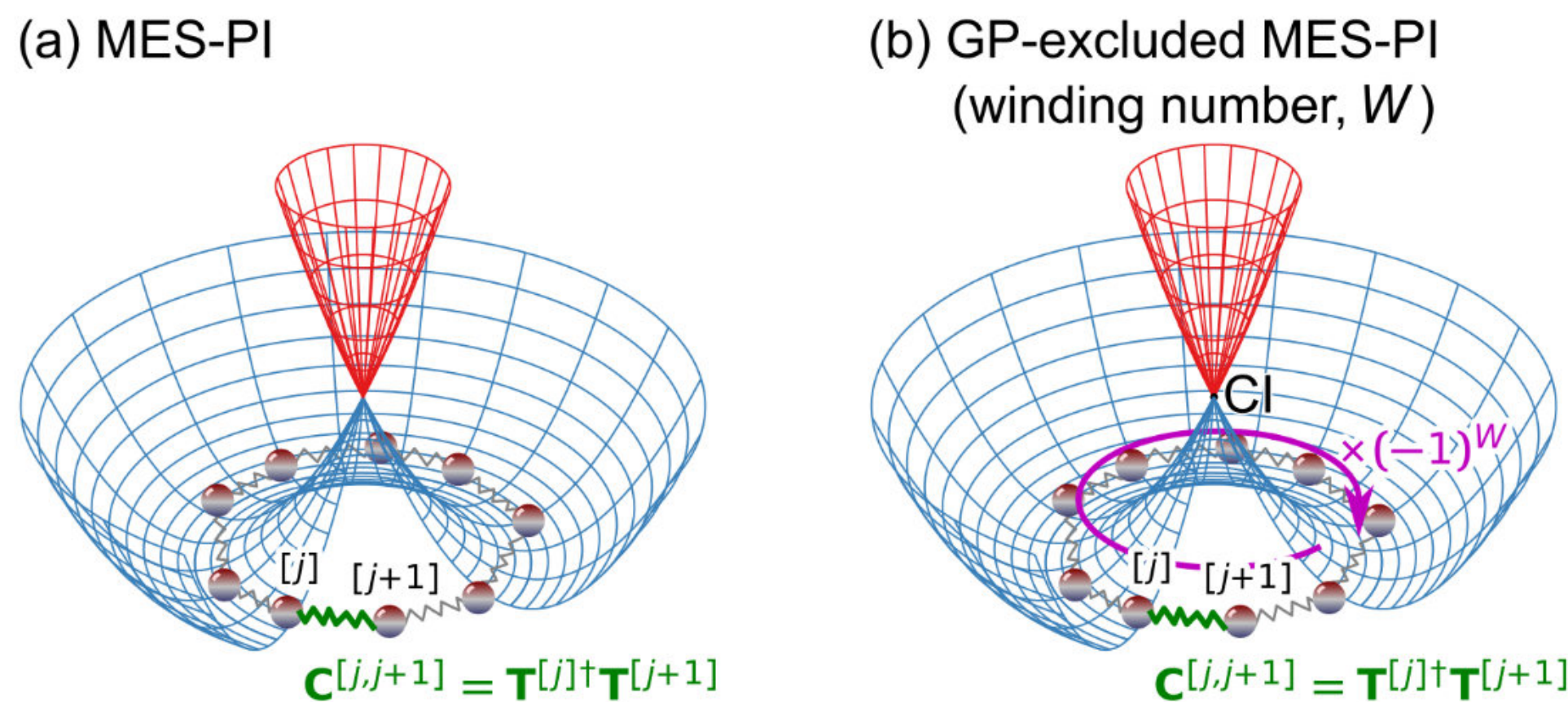

**Figure 2.** Graphical depictions of the MES-PI approach and the artificial GP-excluded MES-PI constructed using the winding-number–induced phase factor, when the adiabatic representation is employed. (a) MES-PI (eq (14)), in which both ground and excited electronic-states (in the adiabatic representation) are considered and full overlap matrices $\left\{\mathbf{C}^{[j,j+1]}\right\}$ between successive beads along the ring polymer are employed. (b) GP-excluded MES-PI with the winding number, $W$ (represented by the magenta arrow, see SI Section S2.5.1 for more details).

**Artificial GP-Excluded MES-PI Formulation.** To verify the role of the GP, we introduce an *ad hoc* winding number of the ring polymer, determined by the topology of the ring polymer relative to the CI. For general systems containing multiple CIs, the corresponding topological phase factor is defined by $\prod_{\alpha} \boldsymbol{\eta}_{\alpha}^{W_{\alpha}}$, where $\boldsymbol{\eta}_{\alpha}$ is the (diagonal) geometric signature matrix for the $\alpha$-th CI, and $W_{\alpha}$ is the winding number with respect to this CI. The winding number $W_{\alpha}$ is obtained using a ray-casting algorithm of computational geometry [50], where $W_{\alpha}$ is determined by counting the flux of paths across an arbitrary ray projecting from the CI on the $g$ - $h$ plane. (See refs [7, 51, 52] on the $g$ - $h$ plane.) While the ray-casting algorithm evaluates both positive and negative winding numbers for the imaginary-time paths, only the magnitude is physically relevant to the topological phase; thus, throughout this paper, $W_\alpha$ denotes the absolute winding number. When the GP is artificially excluded by multiplying the winding-number-induced phase factor in *adiabatic-MES-PI* [24] (eq (14)), the expression of the artificial GP-excluded partition function reads

$$Z^{\text{adia}}_{N_{\text{bead}},\text{GP-excluded}} = \left| \frac{N_{\text{bead}}\mathbf{M}}{2\pi\beta\hbar^2} \right|^{\frac{N_{\text{bead}}}{2}} \int \cdots \int \mathrm{d}\left\{\mathbf{R}^{[j]}\right\} \mathrm{e}^{-\beta U_{\text{spr}}} \mathrm{Tr}_{\text{e}} \left[ \left( \prod_{\alpha} \boldsymbol{\eta}_{\alpha}^{W_\alpha} \right) \prod_{j=1}^{N_{\text{bead}}} \mathrm{e}^{-\frac{\beta}{N_{\text{bead}}}\boldsymbol{\Lambda}^{[j]}} \mathbf{C}^{[j,j+1]} \right] . \quad (15)$$

Equation (15) is a unified expression of the artificial GP-excluded partition function in general MES systems, in which the GP effect is explicitly removed. Contrasting these artificial GP-excluded results (from eq (15)) with those results obtained from the original MES-PI formulation (eq (14)) allows for the precise quantification of topological contributions to the thermodynamic properties in MES systems.

Specifically, for the two-state single-CI Jahn-Teller model discussed in this Perspective, the geometric signature matrix is

$$\boldsymbol{\eta} = \begin{pmatrix} -1 & 0 \\ 0 & -1 \end{pmatrix} , \quad (16)$$

and the corresponding topological phase factor reads

$$\boldsymbol{\eta}^{W} = \begin{pmatrix} -1 & 0 \\ 0 & -1 \end{pmatrix}^{W} = \left(-1\right)^{W} \mathbf{1}_{\text{ele}} \quad (17)$$

with $\mathbf{1}_{\text{ele}}$ the identity matrix in the electronic space. Equation (15) for the GP-excluded artificial partition function is then simplified as

$$Z^{\text{adia}}_{N_{\text{bead}},\text{GP-excluded}} = \left| \frac{N_{\text{bead}}\mathbf{M}}{2\pi\beta\hbar^2} \right|^{\frac{N_{\text{bead}}}{2}} \int \cdots \int \mathrm{d}\left\{\mathbf{R}^{[j]}\right\} \mathrm{e}^{-\beta U_{\text{spr}}} \mathrm{Tr}_{\text{e}} \left[ \left(-1\right)^{W} \prod_{j=1}^{N_{\text{bead}}} \mathrm{e}^{-\frac{\beta}{N_{\text{bead}}}\boldsymbol{\Lambda}^{[j]}} \mathbf{C}^{[j,j+1]} \right] . \quad (18)$$

Thus, the winding-number-induced phase factor can be represented by $(-1)^W$ for this present single-CI model. Figure 2(b) presents a schematic of the artificial GP-excluded MES-PI approach in the adiabatic representation, illustrating eq (18) for the single-CI system. Please see more discussions in SI Section S2.5.1. In MES systems involving multiple CIs, the non-zero elements of diagonal geometric signature matrix $\boldsymbol{\eta}_\alpha$ are $\pm 1$, depending on the nature of the $\alpha$-th CI. In SI Section S2.5.3, we provide an illustrative example of the geometric signature matrices for a three-state system containing three CIs, featuring a combination of Jahn-Teller and pseudo-Jahn-Teller effects [28]. It includes both topologically trivial and nontrivial CIs.

Historically, refs [53] and [54] showed that within the real-time PI formulation for a two-dimensional problem, the analytical propagator can be decomposed into contributions from paths with different winding numbers, which naturally leads to the introduction of a winding-number-induced phase factor $(-1)^W$. While refs [55] and [56] followed refs [53] and [54] in assigning winding numbers to individual real-time paths, no explicit methodology for directly implementing the path integral was provided. In contrast, in imaginary-time PI simulations of liquid $^4$He (bosonic particles) in the superfluid regime, the winding number is explicitly defined by counting the flux of paths across an arbitrary plane[48]. In this case, no winding-number-induced phase factor is required for the investigation of superfluidity. To the best of our knowledge, the imaginary-time PI study presented in this Perspective (and the SI) is the first to explicitly resolve both the winding numbers $\{W_\alpha\}$ and the corresponding topological phase factor $\prod_\alpha \boldsymbol{\eta}_\alpha^{W_\alpha}$ (with the geometric signature matrices $\{\boldsymbol{\eta}_\alpha\}$ for both topologically trivial and nontrivial CIs) for each individual path within the ensemble of the general MES system.

**Role of the GP in MES-PI and in Thermodynamic Properties.** In Figure 3, we consider the MES-PIMD simulation of the Jahn-Teller model (eq (1)) with $c = 1$ r.u. at $\beta = 8$ r.u.. Mapping the reduced units onto the Jahn-Teller system, $Ba_2MgReO_6$, of ref [39], we find that these parameters yield a temperature of approximately 70 K. We employ this case to demonstrate the winding behaviors of the ring polymer with respect to the CI. Figure 3(a-b) shows the probability distribution of the winding number. As the winding number increases, the probability decreases exponentially. In the simulation, the percentage of $W = 0$ is over 60%, that of $W = 1$ is close to 35%, but that of $W \geq 3$ is only around 0.3%. The distribution of the winding number can be rationalized. In the same PIMD simulation, the average "lengths" of the ring polymers with different winding numbers are nearly the same, which are roughly determined by the inverse temperature $\beta$. With an increasing winding number, the effective length per loop is shortened, indicating a more constrained path geometry around the CI. This implies that the ring polymer must sample regions of higher ground-state potential energy surrounding the CI (Figure 1(a)); consequently, the associated Boltzmann weight decreases, causing the probability of such ring polymer configurations to drop exponentially. Odd integers $W$ occur with a proportion larger than $1/3$. It suggests that the GP will have a profound impact on thermodynamic properties. Figure 3(c) depicts the three representative ring polymer configurations for $W = 0, 1,$ and $2$, respectively, from the MES-PIMD simulation. When the positive half of the $x$ axis (the gray solid line) is used as a ray projecting from the CI, the winding number is calculated as the net sum of crossings across the ray, where each crossing is assigned a value of $+1$ or $-1$ based on its direction.

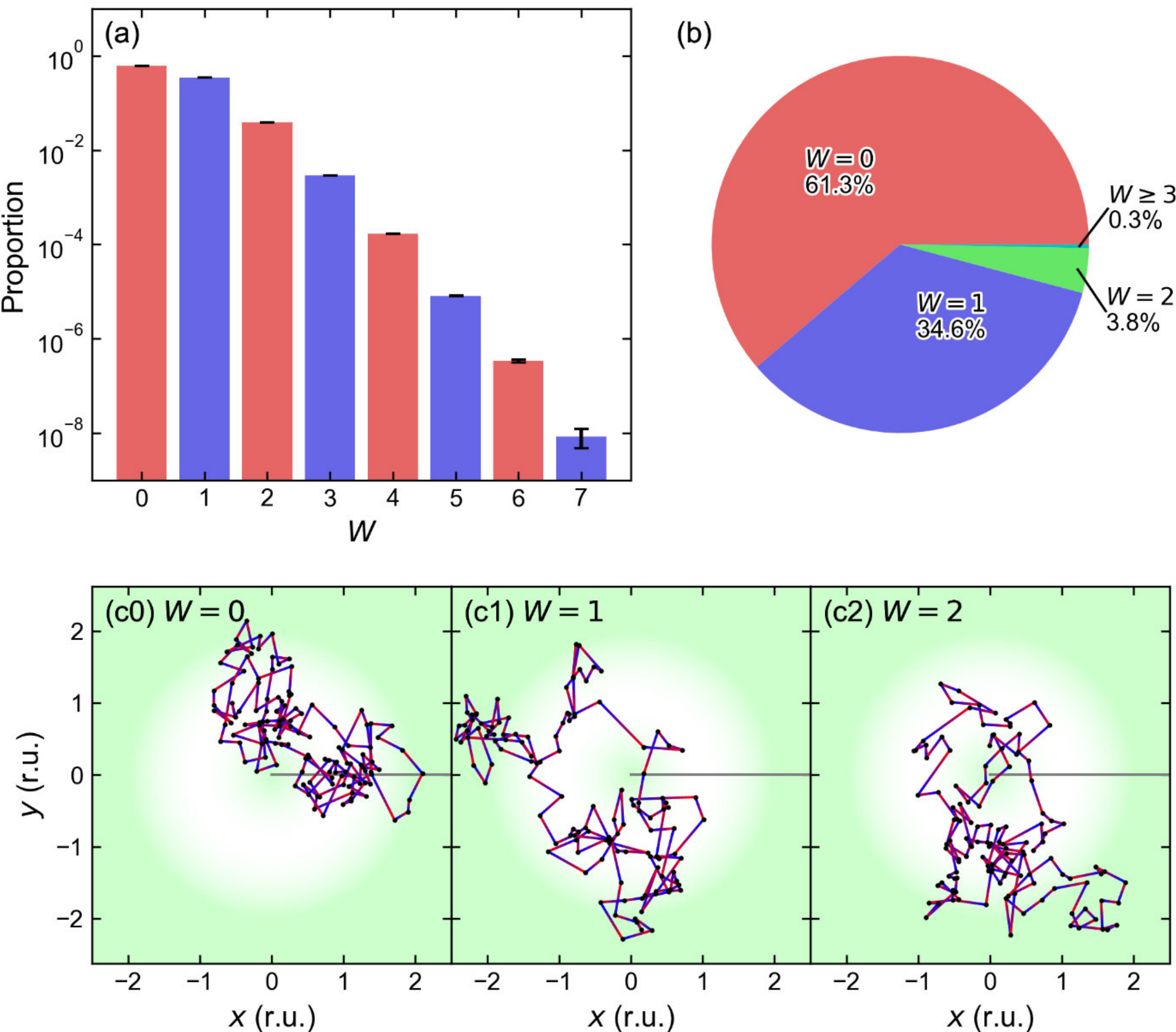

**Figure 3.** Winding numbers of the ring polymer configurations with 128 beads in the MES-PIMD simulation for the Jahn-Teller model (eq (1)) with $c = 1$ r.u. at $\beta = 8$ r.u. . (a) Histogram illustrating the distribution of the winding number. The proportion drops exponentially as $W$ increases. Red bars represent even winding numbers, while blue bars stand for odd winding numbers. (b) Percentage distribution of winding numbers for ring polymer configurations. (c0-c2) Three representative ring polymer configurations extracted from the MES-PIMD simulation, exhibiting winding numbers $W = 0$, $1$, and $2$, respectively. Each link between two successive beads is colored with a gradient (from red to blue) to mark the direction of the imaginary-time.

The positive half of the $x$ axis (solid gray line) is employed as the reference ray for determining the winding number of the ring polymer configuration. The winding number is obtained by counting the net number of links crossing the gray line (i.e., the difference between clockwise and counterclockwise crossings). The background green color represents the PES of the ground electronic-state. Panel (c) illustrates the fractal features of the ring polymer configurations in the imaginary-time path integral.

MES-PIMD simulations [24, 25] are then performed to study thermodynamic properties, including the kinetic energy, potential energy, and heat capacity. The MES-PIMD simulation details, as well as the expressions of the evaluation formulas and estimators, are presented in SI Section S2. Specifically, we introduce an evaluation formula for the heat capacity in MES-PIMD simulations, which includes a double-virial-estimator term [57, 58] and converges significantly faster (as shown in Figure S12).

Figure 4 shows the results as functions of the number of beads $N_{\mathrm{bead}}$ for the Jahn-Teller model with parameter $c = 1$ r.u. at a low temperature ( $\beta = 8$ r.u. ). Figure 4(a)-(c) demonstrate that MES-PIMD yields accurate estimates of all thermodynamic properties (red circles); notably, its converged results (with sufficiently large $N_{\mathrm{bead}}$ ) approach the exact data generated *via* the basis-set-expansion method (gray solid line) for the (correct) GP-included Hamiltonian of eq (9). It verifies that MES-PIMD yields correct physical properties for thermal-equilibrium systems when the GP plays a crucial role. When we artificially exclude the GP in MES-PIMD by using the winding-number-induced phase factor, the results (blue upward triangles) in the limit $N_{\mathrm{bead}} \to \infty$ eventually converge to the values produced by the basis-set-expansion method (gray dashed line) for the (artificial) GP-excluded Hamiltonian of eq (4). It indicates that the GP-excluded MES-PIMD approach with the winding-number-induced phase factor offers a practical tool for investigating GP-excluded thermodynamic properties.

Interestingly, as shown in Figure 4(a)-(c) , the GP-excluded MES-PIMD approach by brute-forcely using the winding number converges much more slowly. The results (blue upward triangles) are not converged yet even with $2^{12}=4096$ beads. The fitting of the data reveals that the asymptotic behavior of property $B_{N_{\text{bead}}}$ obtained from the GP-excluded MES-PI simulation follows $B_{N_{\text{bead}}}=B_{\text{ext}}+b/\sqrt{N_{\text{bead}}}$ when $N_{\text{bead}}$ is sufficiently large. The extrapolation values of $B_{\text{ext}}$ for the kinetic energy, potential energy, and heat capacity all agree with the results provided by the basis-set-expansion method for the GP-excluded Hamiltonian of eq (4) (see SI Section S2.7.3 for more details). In contrast, the bias in a conventional Trotter-splitting PI simulation decreases as $1/N_{\text{bead}}^2$ in the limit $N_{\text{bead}}\to\infty$. That is, the asymptotic behavior is $B_{N_{\text{bead}}}=B_{\text{ext}}+\tilde{b}/N_{\text{bead}}^2$ when the Trotter decomposition is used in MES-PI [44, 45]. The significant difference in the asymptotic behavior as the number of beads, $N_{\text{bead}}$, increases reveals the profound GP effect on the numerical convergence of PI simulations of nonadiabatic systems. In contrast to the GP-included case's smooth behavior at the origin (Figure S2 of the Supporting Information), excluding the GP makes the system evolve on the adiabatic cone without the compensating vector potential, effectively creating a singularity (cusp) in the action at the CI point (as shown by the singular behavior of the ground-state wavefunction, SI Figure S3), which destroys the second-order error cancellation in the Trotter decomposition. It is expected that the degradation always arises even when a higher-order decomposition [59] is employed in the brute-force GP-excluded MES-PI approach. As described in SI Section S2.5.2, an efficient algorithm that absorbs the GP into the spring potential $U_{\mathrm{spr}}$ (GPA-SP) is proposed for the artificial, GP-excluded MES-PI approach, which restores the bias convergence at $O\left(1/N_{\mathrm{bead}}^{2}\right)$ when the Trotter decomposition is used.

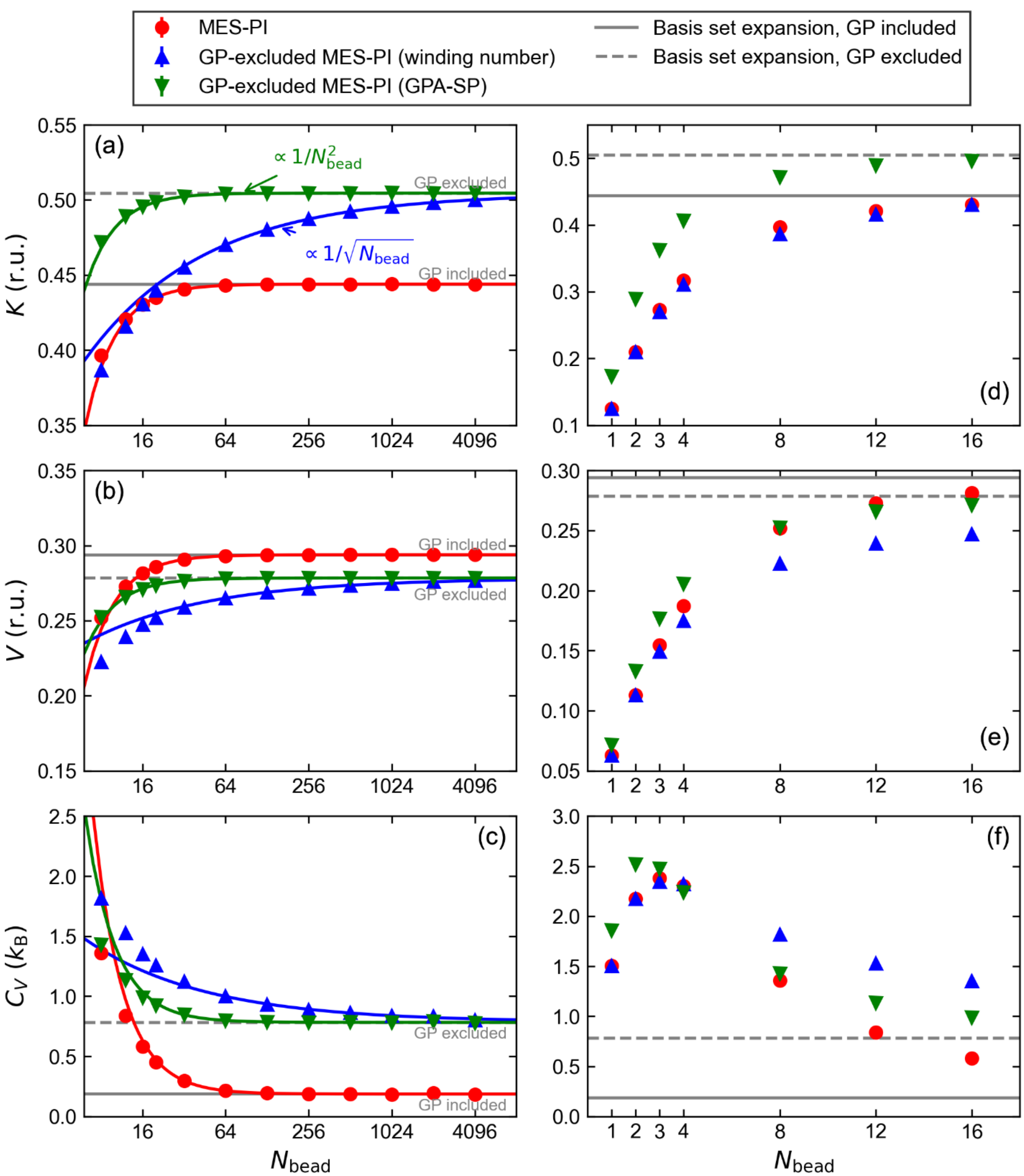

**Figure 4.** MES-PIMD simulation results of the Jahn-Teller model (eq (1)) with $c = 1$ r.u. at $\beta = 8$ r.u.. Both GP-included and GP-excluded cases are studied. (a, d) Average kinetic energy (with the virial estimator in eq (S54)); (b, e) Average potential energy; and (c, f) Heat capacity (evaluated by eq (S60), where a double-virial-estimator term and two virial-estimator terms are involved). Red Circles: GP-included case; Blue Upward Triangles: GP-excluded case where the winding-number-induced phase factor along the ring polymer directly cancels the GP; Green Downward Triangles: GP-excluded case with the GPA-SP algorithm (SI Section S2.5.2). In each case, the PIMD results versus the number of beads are fitted with the corresponding asymptotic formula in the regime where $N_{\text{bead}}$ is sufficiently large. Each asymptotic formula is represented by the solid line in the same color. Gray Solid Lines: Benchmarks of the GP-included case by using the basis-set-expansion method to directly solve the time-independent Schrödinger equation for the Hamiltonian of eq (9); Gray Dashed Lines: Benchmarks of the GP-excluded case by using the basis-set-expansion method to calculate the eigensystems of the Hamiltonian of eq (4).

Figure 4 (d)-(f) demonstrate the results obtained when $N_{\text{bead}}$, the number of beads, is relatively small. In the imaginary-time PI formulation, the $N_{\text{bead}} = 1$ case represents "the classical-nuclei limit". Because the GP is a pure quantum effect, MES-PIMD and the GP-excluded MES-PIMD approach by brute-forcely using the winding number lead to the same results in the classical-nuclei limit. As it is impossible for a ring polymer of only two beads to wind around the CI, the winding number is always 0 and the $N_{\text{bead}} = 2$ case also captures no GP effect. It is confirmed in Figure 4 (d)-(f) of Figure 4 where red circles and blue upward triangles are identical results for all thermodynamic properties when $N_{\text{bead}} = 1$ or $N_{\text{bead}} = 2$. In contrast, since the GPA-SP algorithm introduces the quantum "erf" hole created by the GP even when $N_{\text{bead}} = 1$, its results (green downward triangles) do not simply approach the classical ones. Figure 4 shows that it is crucial to use sufficiently large $N_{\text{bead}}$ to study the GP effect in thermodynamic properties. For instance, the numerical convergence behavior of the heat capacity as a function of $N_{\text{bead}}$ is not monotonic, and the relatively small number of beads can lead to even qualitatively wrong conclusions. As demonstrated in Figure 4(f), when $N_{\text{bead}}$ is between 2 and 4, the "quantum" results of the heat capacity seem greater than the value of the classical-nuclei limit; therefore, one might conclude that the GP effect yields a larger value of the heat capacity. This is, however, misleading, because the converged quantum result of the heat capacity is lower than its classical-nuclei limit. As the GP effect is coupled with other nuclear quantum effects, rigorous results for thermodynamic properties should be obtained in the limit $N_{\text{bead}} \to \infty$, and it is important to use converged PI results to conduct the investigation.

Note that the artificial, GP-excluded MES-PI approach is employed *only* for comparison with the standard MES-PI to demonstrate the effect of the GP on thermodynamics. The MES-PI formulation naturally accounts for the GP, yielding correct converged results for thermodynamic properties. The winding-number-induced phase factor should *never* be incorporated in any MES-PI simulations of real systems.

**Single-Electronic-State Limit**. Finally, we examine the single-electronic-state (SES) limit, commonly used in regimes where excited electronic-states are relatively high in energy and the temperature is relatively low (see, e.g., refs. [18, 29, 60, 61] and references therein). A more compact form for both eq (4) and eq (9) reads

$$\hat{\mathbf{H}}_{\text{adia}} = \frac{1}{2}\left(\hat{\mathbf{P}}_{\text{can}} + \mathbf{A}_{\text{GP}}\left(\hat{\mathbf{R}}\right) - \mathrm{i}\mathbf{d}\left(\hat{\mathbf{R}}\right)\right)\cdot\left(\hat{\mathbf{P}}_{\text{can}} + \mathbf{A}_{\text{GP}}\left(\hat{\mathbf{R}}\right) - \mathrm{i}\mathbf{d}\left(\hat{\mathbf{R}}\right)\right) + \hat{\mathbf{\Lambda}}$$
$$= -\frac{1}{2}\left[\frac{\partial^2}{\partial r^2} + \frac{1}{r}\frac{\partial}{\partial r} + \frac{1}{r^2}\left(\frac{\partial}{\partial \theta} + A_{\text{GP}}^{(\theta)} + \mathbf{d}^{(\theta)}\right)^2\right] + \frac{1}{2}\begin{pmatrix} (r-c)^2 & 0 \\ 0 & (r+c)^2 \end{pmatrix}. \tag{19}$$

Here, both $\mathbf{A}_{\text{GP}}$ and $\mathbf{d}$ are matrices, in which each element is a vector in the nuclear space. The vector potential corresponding to the GP is $\mathbf{A}_{\text{GP}} = \frac{-\mathrm{i}A_{\text{GP}}^{(\theta)}}{r}\begin{pmatrix} \mathbf{e}_\theta & 0 \\ 0 & \mathbf{e}_\theta \end{pmatrix}$, where $\mathbf{e}_\theta$ is the unit vector in $\theta$ direction, which is $\left(\frac{-y}{\sqrt{x^2+y^2}}, \frac{x}{\sqrt{x^2+y^2}}\right)^T$ in Cartesian coordinates. $A_{\text{GP}}^{(\theta)} = 0$ for the GP-excluded case, while $A_{\text{GP}}^{(\theta)} = \mathrm{i}/2$ for the GP-included case. In either case, the nonadiabatic coupling vector is $\mathbf{d} = \frac{\mathbf{d}^{(\theta)}}{r}\otimes\mathbf{e}_\theta = \frac{\mathrm{i}\hat{\boldsymbol{\sigma}}_y}{2r}\otimes\mathbf{e}_\theta = \frac{1}{2r}\begin{pmatrix} 0 & \mathbf{e}_\theta \\ -\mathbf{e}_\theta & 0 \end{pmatrix}$. When only the ground electronic-state is considered, the Born-Huang adiabatic (BHA) approximation [26, 62-64] to eq (19) generates the Hamiltonian

$$\hat{H}_{\text{BHA}}^{\text{GP}} = -\frac{1}{2}\left[\frac{\partial^2}{\partial r^2} + \frac{1}{r}\frac{\partial}{\partial r} + \frac{1}{r^2}\left(\frac{\partial}{\partial \theta} + \frac{\mathrm{i}}{2}\right)^2\right] + \frac{1}{8r^2} + \frac{1}{2}(r-c)^2 \tag{20}$$

for the GP-included case, and the Hamiltonian

$$\hat{H}_{\text{BHA}}^{\text{BF}} = -\frac{1}{2}\left[\frac{\partial^2}{\partial r^2} + \frac{1}{r}\frac{\partial}{\partial r} + \frac{1}{r^2}\frac{\partial^2}{\partial \theta^2}\right] + \frac{1}{8r^2} + \frac{1}{2}(r-c)^2 \tag{21}$$

for the GP-excluded case. Similarly, the BO approximation [26, 62-64] to eq (19) produces the Hamiltonian

$$\hat{H}_{\text{BO}}^{\text{GP}} = -\frac{1}{2}\left[\frac{\partial^2}{\partial r^2} + \frac{1}{r}\frac{\partial}{\partial r} + \frac{1}{r^2}\left(\frac{\partial}{\partial \theta} + \frac{\mathrm{i}}{2}\right)^2\right] + \frac{1}{2}(r-c)^2 \tag{22}$$

for the GP-included case, and the Hamiltonian

$$\hat{H}_{\mathrm{BO}}^{\mathrm{BF}} = -\frac{1}{2}\left[\frac{\partial^2}{\partial r^2} + \frac{1}{r}\frac{\partial}{\partial r} + \frac{1}{r^2}\frac{\partial^2}{\partial \theta^2}\right] + \frac{1}{2}(r-c)^2 \tag{23}$$

for the GP-excluded case. BHA differs from BO through the inclusion of the DBOC term, $1/8r^2$. The basis-set-expansion method is used to compute vibrational eigenstates and then thermodynamic properties for each SES Hamiltonian of eqs (20)-(23). In this approach, the eigenstates of the corresponding nuclear kinetic energy operator are employed as the basis set for solving the time-independent Schrödinger equation. This choice ensures efficient convergence of thermodynamic quantities because, near the CI point $r = 0$ (i.e., the origin), the nuclear kinetic energy operator with radial-derivatives and $1/r^2$ terms dominates the short-range behavior of the nuclear wavefunctions, whereas the potential energy remains finite. The eigenfunctions of the kinetic-energy operator, whose radial parts are given by Bessel functions, are therefore well suited to capture the singular short-range behavior—the cusp-like structure, the divergent derivative and the strictly vanishing probability at the origin—of the ground-state wavefunctions of the GP-included BO Hamiltonian (eq (22)), the GP-excluded BHA Hamiltonian (eq (21)), and the GP-included BHA Hamiltonian (eq (20)), as illustrated in Figures S4 and S5. In contrast, the ground-state wavefunction of the GP-excluded BO Hamiltonian (eq (23)) remains smooth at the origin. Additional discussions and computational details are provided in SI Section S1.

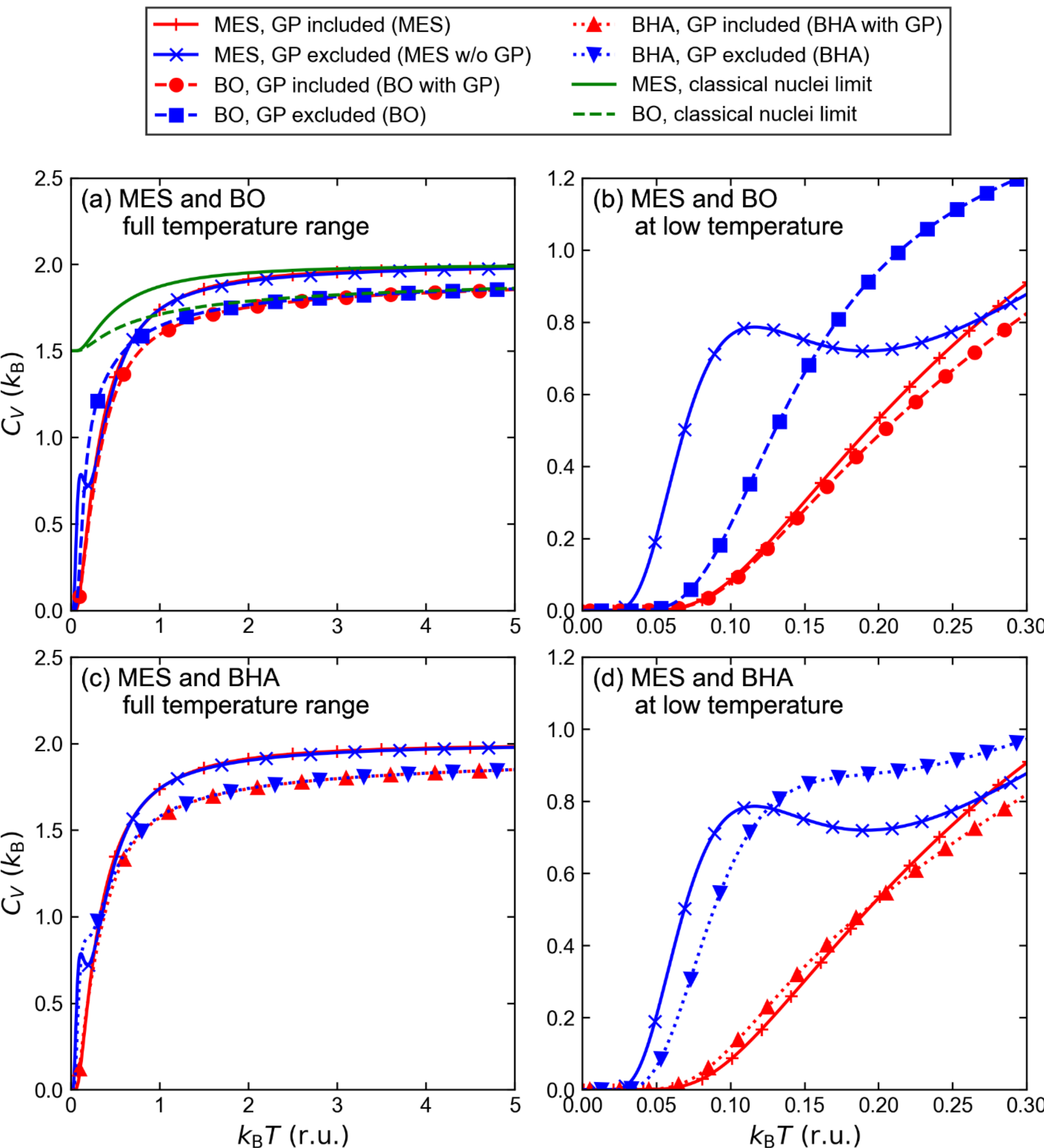

**Figure 5.** Temperature dependence of the heat capacity ($C_V$) of the Jahn-Teller model (eq (1)) with parameter $c = 1\,\text{r.u.}$, obtained by the basis-set-expansion method. (a, c): Full temperature regime; (b, d): low-temperature regime. Panels (a) and (b) compare the results of MES (eqs (4)

and (9)) to those produced by the BO approximation (eqs (22) and (23)). Panels (c) and (d) compare the results of MES to those obtained by the BHA approximation (eqs (20) and (21)). GP-included cases are presented in red, GP-excluded cases in blue, and classical-nuclei-limit cases in green. Solid lines represent results with two electronic states, dashed lines indicate results with the BO approximation, and dotted lines correspond to results with the BHA approximation.

Figure 5 presents the heat capacity results for the Jahn-Teller model with parameter $c = 1\,\text{r.u.}$, where the potential energy of the CI point is relatively small. In comparison, Figure 6 shows the corresponding results for the same model with $c = 3\,\text{r.u.}$, where the potential energy of the CI point is relatively large. As shown in Figure 5(a), (c) as well as in Figure 6(a), (c), the SES limit breaks down in the high-temperature regime, regardless of whether BO or BHA is used. At high temperatures, the MES heat capacity is greater than the SES heat capacity. In the high-temperature regime, both the GP effect and other nuclear quantum effects are negligible (in the heat capacity), so the MES classical-nuclei limit becomes valid, as illustrated in Figure 5(a) and Figure 6(a). The MES classical-nuclei-limit loses its validity in the low-temperature regime, where nuclear quantum effects become important. Figure 5(a) and Figure 6(a) show that the classical-nuclei limit leads to finite heat capacity at $0\,\text{K}$, which is unphysical. At low temperatures, the inclusion of the GP is essential for a reasonable description of the heat capacity (and other thermodynamic properties), even when the nuclear DOFs are treated quantum-mechanically. Figure 5(b), (d), Figure 6(b), (d), and Figure 1(d) demonstrate that the GP-excluded MES Hamiltonian in eq (4) produces an artificial peak in the heat capacity in the low-temperature regime.

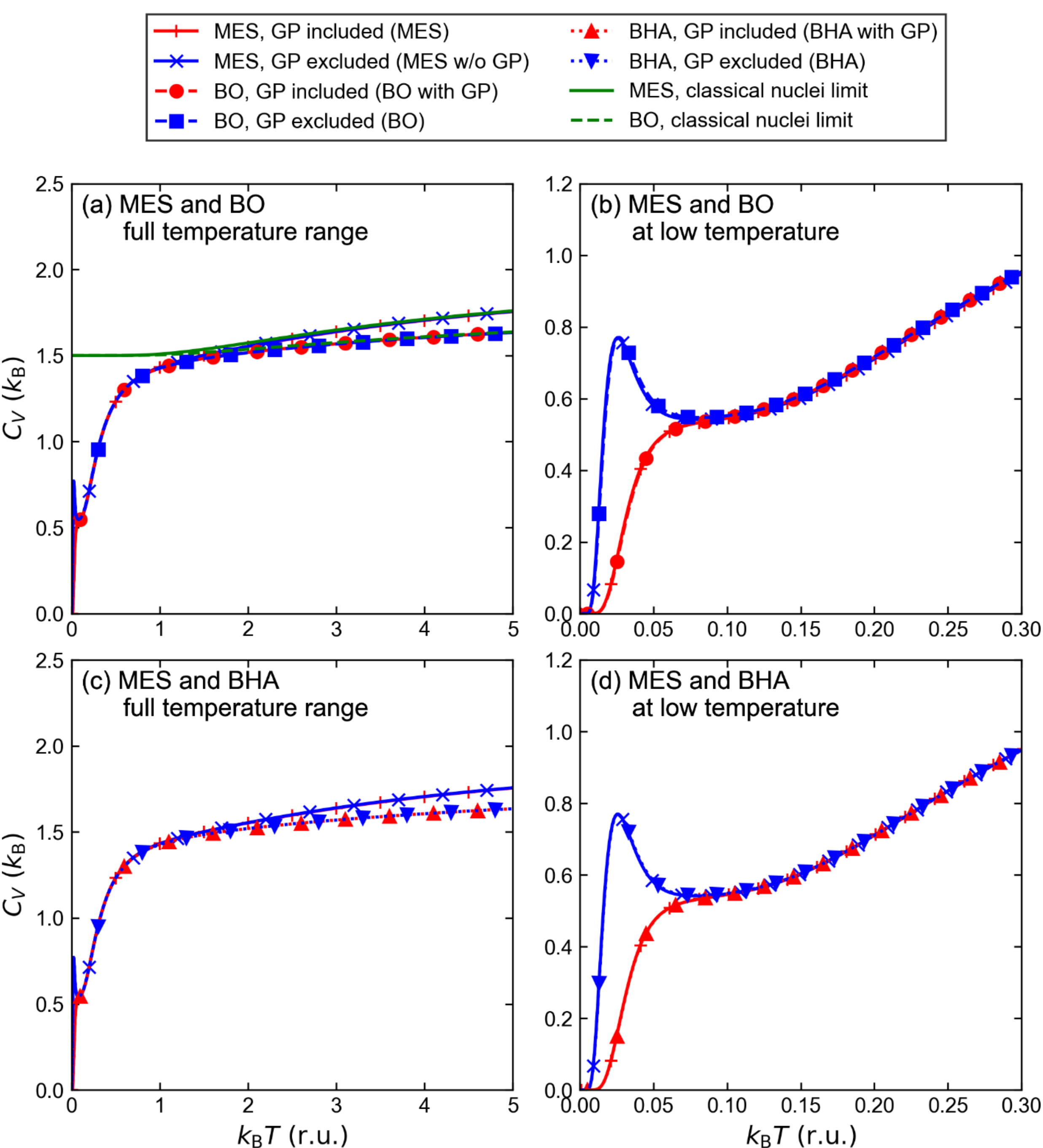

**Figure 6.** Same as Figure 5, but for parameter $c = 3$ r.u..

Figure 5(b),(d) illustrate that, for parameter $c = 1$ r.u. where the adiabatic potential-energy gap is relatively small, the BO and BHA results deviate considerably from the MES benchmarks. The discrepancy demonstrates that the MES formulation is essential in regimes where simplified corrections to BO and BHA fail to capture all topological and nonadiabatic effects. In contrast, Figure 6(b), (d) demonstrate that, for parameter $c = 3$ r.u., the BHA results show excellent agreement with the MES data, while the BO approximation exhibits only marginal deviations. The GP-included Hamiltonian in the SES limit accurately describes the low-temperature regime for systems with a relatively large potential-energy gap between the ground and excited adiabatic PESs. A more complete comparison between the MES results and those in the SES limit with respect to the value of parameter $c$ is shown in Movie S1 of the Supporting Information.

**Single-Electronic-State Path Integral Schemes.** In the SES limit, we further consider three imaginary-time PI schemes in the adiabatic representation that are practically useful for general molecular systems. The first PI scheme, denoted by BO-PI, involves the BO approximation, where the expression of the partition function is

$$Z_{N_{\mathrm{bead}}}^{\mathrm{BO}} = \left| \frac{N_{\mathrm{bead}} \mathbf{M}}{2\pi\beta\hbar^2} \right|^{\frac{N_{\mathrm{bead}}}{2}} \int \cdots \int \mathrm{d}\left\{ \mathbf{R}^{[j]} \right\} \mathrm{e}^{-\beta U_{\mathrm{spr}}} \exp\left[ \sum_{j=1}^{N_{\mathrm{bead}}} - \frac{\beta}{N_{\mathrm{bead}}} \lambda_1^{[j]} \right], \tag{24}$$

where $\lambda_1^{[j]} = \lambda_1\left(\mathbf{R}^{[j]}\right)$ is the ground electronic-state potential-energy for nuclear coordinates $\mathbf{R}^{[j]}$. Equation (24) is the most conventional PI formula in the SES limit. Specifically, in the two-dimensional model, the electronic ground state potential energy $\lambda_1\left(\mathbf{R}\right) = \frac{1}{2}\left(c - r\right)^2$ is the first diagonal element of the adiabatic PES matrix (eq (5)).

The second PI scheme, denoted by BHA-PI, employs the BHA approximation, where the DBOC term for the ground electronic-state remains when all excited-state-related terms are omitted. Its expression of the partition function then reads

$$Z_{N_{\text{bead}}}^{\text{BHA}} = \left|\frac{N_{\text{bead}}\mathbf{M}}{2\pi\beta\hbar^2}\right|^{\frac{N_{\text{bead}}}{2}} \int\cdots\int \mathrm{d}\left\{\mathbf{R}^{[j]}\right\} \mathrm{e}^{-\beta U_{\text{spr}}} \exp\left[\sum_{j=1}^{N_{\text{bead}}} -\frac{\beta}{N_{\text{bead}}}\lambda_1^{[j]}\right] \exp\left[\sum_{j=1}^{N_{\text{bead}}} -\frac{\beta}{N_{\text{bead}}} K_{\text{DBOC}}^{[j]}\right], \quad (25)$$

where

$$K_{\text{DBOC}} = -\sum_K \frac{\hbar^2}{2M_K}\left\langle \phi_1 \middle| \frac{\partial^2\phi_1}{\partial R_K^2}\right\rangle = \sum_K \frac{\hbar^2}{2M_K}\left\langle \frac{\partial\phi_1}{\partial R_K} \middle| \frac{\partial\phi_1}{\partial R_K}\right\rangle. \quad (26)$$

When the DBOC term is negligible, BHA-PI becomes BO-PI.

Both BO-PI and BHA-PI intrinsically neglect the GP effect. However, the GP can be incorporated *a posteriori* by introducing a phase factor $\eta_\alpha^{W_\alpha}$ determined by the ring polymer's winding number ($W_\alpha$) with respect to the $\alpha$ -th CI. Here, the geometric signature number $\eta_\alpha$ is the ground-electronic-state diagonal matrix element of the geometric signature matrix $\boldsymbol{\eta}_\alpha$ of the $\alpha$ -th CI, which takes the value of $-1$ or 1, depending on its topological nature. That is, the expression of the partition function of GP-included BO-PI reads

$$Z_{N_{\text{bead}},\text{GP-included}}^{\text{BO}} = \left|\frac{N_{\text{bead}}\mathbf{M}}{2\pi\beta\hbar^2}\right|^{\frac{N_{\text{bead}}}{2}} \int\cdots\int \mathrm{d}\left\{\mathbf{R}^{[j]}\right\} \mathrm{e}^{-\beta U_{\text{spr}}} \exp\left[\sum_{j=1}^{N_{\text{bead}}} -\frac{\beta}{N_{\text{bead}}}\lambda_1^{[j]}\right]\left(\prod_\alpha \eta_\alpha^{W_\alpha}\right), \quad (27)$$

and that of GP-included BHA-PI is

$$\begin{aligned} Z_{N_{\text{bead}},\text{GP-included}}^{\text{BHA}} = \left|\frac{N_{\text{bead}}\mathbf{M}}{2\pi\beta\hbar^2}\right|^{\frac{N_{\text{bead}}}{2}} \int\cdots\int \mathrm{d}\left\{\mathbf{R}^{[j]}\right\} \mathrm{e}^{-\beta U_{\text{spr}}} &\exp\left[\sum_{j=1}^{N_{\text{bead}}} -\frac{\beta}{N_{\text{bead}}}\lambda_1^{[j]}\right] \\ &\times\exp\left[\sum_{j=1}^{N_{\text{bead}}} -\frac{\beta}{N_{\text{bead}}} K_{\text{DBOC}}^{[j]}\right]\left(\prod_\alpha \eta_\alpha^{W_\alpha}\right) \end{aligned}. \quad (28)$$

For the present single-CI model, we have $\prod_\alpha \eta_\alpha^{W_\alpha} = (-1)^W$ in eq (27) and eq (28). We note that, when the SES limit is reasonable, the diagonal elements involving only the ground electronic-state in the matrix product $\prod_{j=1}^{N_{\text{bead}}} \mathrm{e}^{-\frac{\beta}{N_{\text{bead}}}\mathbf{\Lambda}^{[j]}} \mathbf{C}^{[j,j+1]}$ in eq (15) dominates over any other elements associated with excited electronic-states. That is, only the ground electronic-state should be considered. Consequently, the winding-number-induced phase factor $\prod_{\alpha} \boldsymbol{\eta}_{\alpha}^{W_\alpha}$ effectively reduces to its ground electronic-state diagonal element, that is, $\prod_{\alpha} \eta_{\alpha}^{W_\alpha} \equiv \prod_{\alpha} \left(\boldsymbol{\eta}_{\alpha}\right)_{1,1}^{W_\alpha}$. Figure 7(a) provides a graphic representation of BO-PI (or BHA-PI), while Figure 7(b) illustrates GP-included BO-PI (or BHA-PI) with a winding-number-induced phase factor for the single-CI system. The GPA-SP algorithm (proposed in SI Section S2.5.2), originally developed for efficient GP-excluded MES-PI simulations, is expected to also improve the numerical convergence behavior of GP-included BO-PI (or BHA-PI).

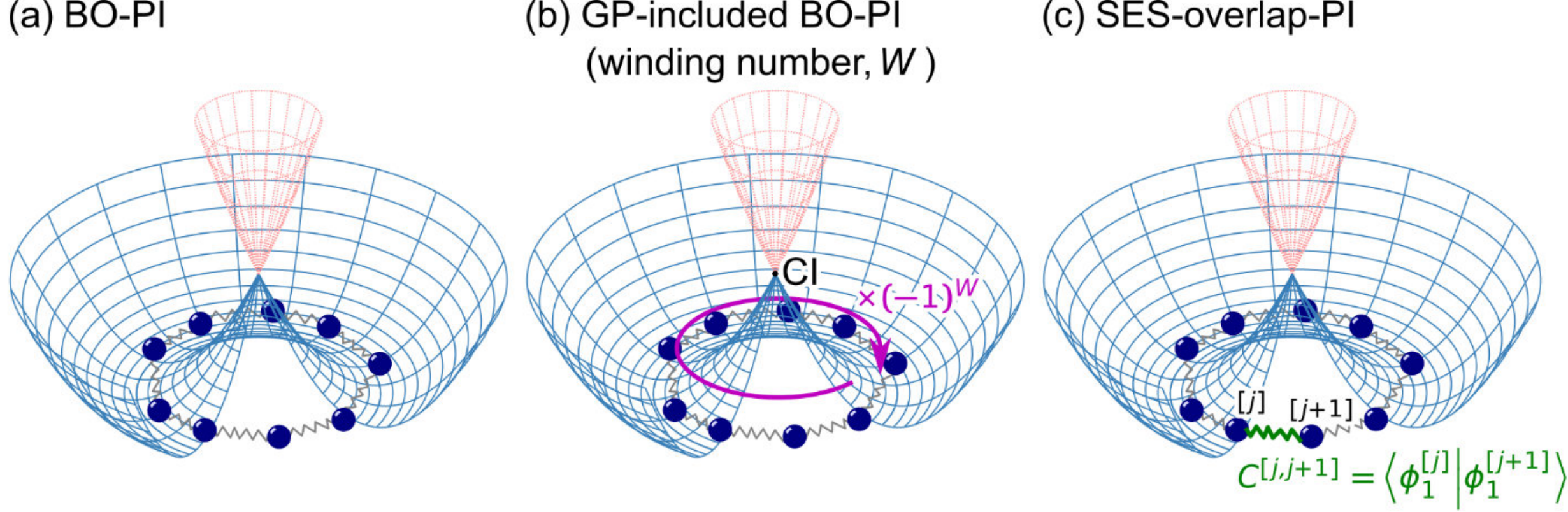

**Figure 7.** Schematic representations of the three different imaginary-time PI approaches in the SES limit. (a) Conventional BO-PI (eq (24)). (b) GP-included BO-PI (eq (27)) with the geometric phase obtained from the winding number $W$ (represented by the magenta arrow). Its numerical convergence can be considerably improved by the GPA-SP algorithm (see SI Section S2.5.2 for more details). (c) SES-overlap-PI (eq (29)), which includes the product of the ground-state

electronic wavefunction overlaps $\{C^{[j,j+1]}\}$ between successive beads along the ring polymer. Panels (a) and (b) also depict the graphic representations of BHA-PI and GP-included BHA-PI, respectively, with the DBOC term included in eq (25).

The third PI scheme is directly derived from the MES-PI approach. When only the ground electronic-state is included in eq (14), the partition function expression becomes

$$Z_{N_{\text{bead}}}^{\text{overlap}} = \left| \frac{N_{\text{bead}} \mathbf{M}}{2\pi\beta\hbar^2} \right|^{\frac{N_{\text{bead}}}{2}} \int \cdots \int \mathrm{d}\left\{ \mathbf{R}^{[j]} \right\} \mathrm{e}^{-\beta U_{\text{spr}}} \exp\left[ \sum_{j=1}^{N_{\text{bead}}} -\frac{\beta}{N_{\text{bead}}} \lambda_1^{[j]} \right] \prod_{j=1}^{N_{\text{bead}}} C^{[j,j+1]}, \tag{29}$$

where $C^{[j,j+1]} = \langle \phi_1^{[j]} | \phi_1^{[j+1]} \rangle = \langle \phi_1 \left( \mathbf{R}^{[j]} \right) | \phi_1 \left( \mathbf{R}^{[j+1]} \right) \rangle$ is the overlap of the ground electronic-states of two successive beads along the ring polymer (Figure 7(c)). This PI scheme is referred to as SES-overlap-PI.

As described in SI Section S2.4, the SES-PIMD simulations of all three schemes require only the ground electronic-state PES and its first-order nuclear derivatives. In addition, to evaluate the weighted terms, BHA-PI needs the DBOC term for the ground electronic-state, and SES-overlap-PI requires the overlaps of ground-state electronic wavefunctions of successive beads along the ring polymer. The SES-overlap-PI scheme naturally captures the GP by $\prod_{j=1}^{N_{\text{bead}}} C^{[j,j+1]}$, the product of the overlaps along the ring polymer. Figure 7(c) presents a graphic representation of SES-overlap-PI. Similar to our previous discussion on MES-PI, inclusion of the phase factor arising from the winding number artificially cancels the GP inherent in SES-overlap-PI. That is, the expression of the partition function of artificial GP-excluded SES-overlap-PI reads

$$Z^{\text{overlap}}_{N_{\text{bead}},\text{GP-excluded}} = \left| \frac{N_{\text{bead}}\mathbf{M}}{2\pi\beta\hbar^2} \right|^{\frac{N_{\text{bead}}}{2}} \int\cdots\int \mathrm{d}\left\{\mathbf{R}^{[j]}\right\} \mathrm{e}^{-\beta U_{\text{spr}}} \exp\left[ \sum_{j=1}^{N_{\text{bead}}} -\frac{\beta}{N_{\text{bead}}} \lambda_1^{[j]} \right] \times \left( \prod_{j=1}^{N_{\text{bead}}} C^{[j,j+1]} \right) \left( \prod_{\alpha} \eta_{\alpha}^{W_{\alpha}} \right) . \tag{30}$$

Specifically, when we consider the present single-CI model, we have $\prod_{\alpha} \eta_{\alpha}^{W_{\alpha}} = \left(-1\right)^{W}$ in eq (30).

Figure 8 and Figure 9 show the SES-PIMD results for the three schemes as functions of the number of beads $N_{\text{bead}}$ for $c = 1\,\text{r.u.}$ at $\beta = 8\,\text{r.u.}$. Converged results are obtained when $N_{\text{bead}}$ is sufficiently large. As shown in Figure 8, converged BO-PI and BHA-PI results agree well with the benchmarks (produced by the basis-set-expansion method) for the GP-excluded BO Hamiltonian and GP-excluded BHA Hamiltonian, respectively. When we incorporate the winding-number-induced phase factor into BO-PI and BHA-PI, their converged results match the benchmarks for the GP-included BO Hamiltonian and GP-included BHA Hamiltonian, respectively. Figure 9 shows that in the limit $N_{\text{bead}} \to \infty$, the SES-overlap-PI scheme reproduces the benchmarks for the GP-included BHA Hamiltonian. We show in SI Section S2.4.4 that BHA-PI with the winding-number-induced phase factor (i.e., GP-included BHA-PI) and SES-overlap-PI converge to the same thermodynamic results.

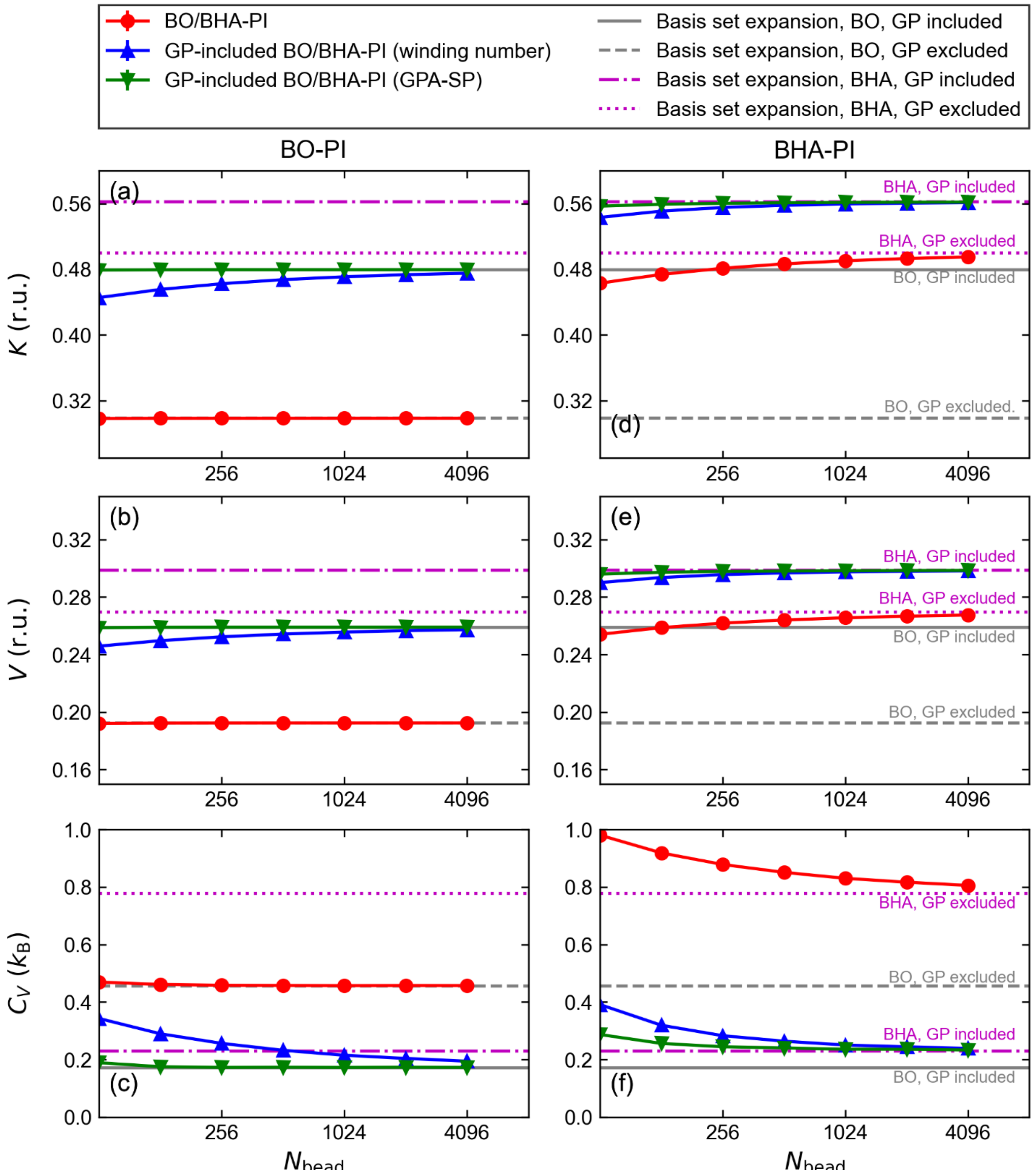

**Figure 8.** BO-PIMD and BHA-PIMD results of the Jahn-Teller model (eq (1)) with $c = 1$ r.u. at $\beta = 8$ r.u.. (a-c): Born-Oppenheimer PI (eq (24)); (d-f): Born-Huang-adiabatic PI (eq (25)). (a, d): Average kinetic energy (with the virial kinetic energy estimator of eq (S54)); (b, e): Average potential energy; (c, f): Heat capacity (with the evaluation formula of eq (S60)). Red circles: no winding number is introduced. Blue upward triangles: GP-included cases where the winding number of the ring polymer is counted. Green downward triangles: GP-included cases with the GPA-SP algorithm (SI Section S2.5.2). Gray solid lines and gray dashed lines indicate benchmarks of the GP-included (eq (22)) and GP-excluded (eq (23)) Hamiltonians under the BO approximation, respectively, calculated *via* the basis-set-expansion method for the time-independent Schrödinger equation. Magenta dash-dotted lines and magenta dotted lines represent benchmarks of the GP-included (eq (20)) and GP-excluded (eq (21)) Hamiltonians under the BHA approximation, respectively, obtained using the basis-set-expansion method for the time-independent Schrödinger equation. The corresponding MES-PIMD results of the same model are shown in Figure 4.

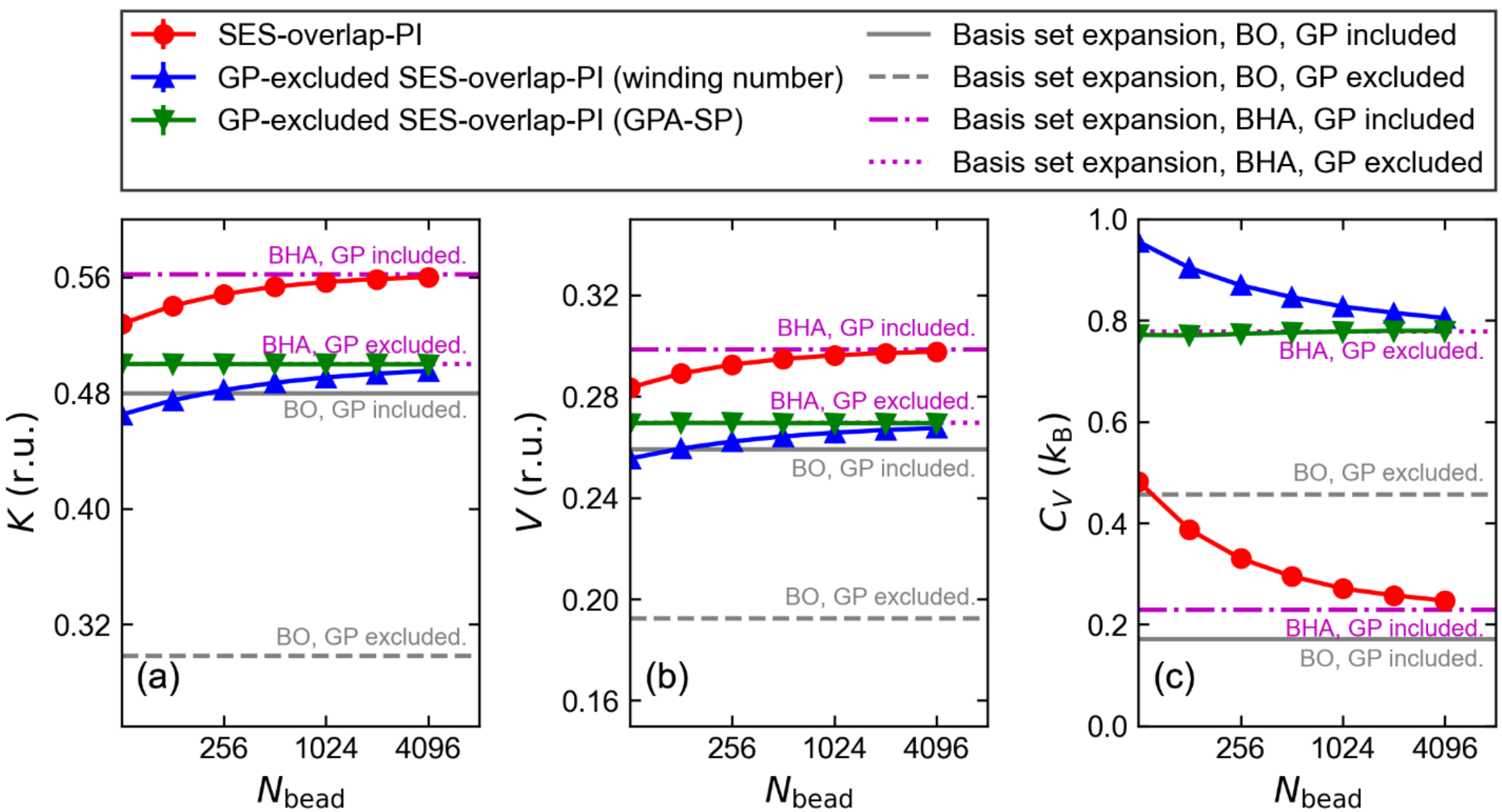

**Figure 9.** SES-overlap-PIMD result of the Jahn-Teller model (eq (1)) with $c = 1$ r.u. at $\beta = 8$ r.u.. (a): Average kinetic energy (with the virial kinetic energy estimator of eq (S54)); (b): Average potential energy; (c): Heat capacity (with the evaluation formula of eq (S60)). Red circles: no winding number is introduced. Blue upward triangles: GP-excluded cases where the winding number of the ring polymer is counted. Green downward triangles: GP-excluded cases with the GPA-SP algorithm (SI Section S2.5.2). Gray solid lines and gray dashed lines indicate benchmarks of the GP-included (eq (22)) and GP-excluded (eq (23)) Hamiltonians under the BO approximation, respectively, calculated *via* the basis-set-expansion method for the time-independent Schrödinger equation. Magenta dash-dotted and magenta dotted lines represent benchmarks of the GP-included (eq (20)) and GP-excluded (eq (21)) Hamiltonians under the BHA approximation, respectively, obtained using the basis-set-expansion method for the time-independent Schrödinger equation. Figure 9 shows that SES-overlap-PI reproduces the benchmarks of the GP-included BHA Hamiltonian (eq (20)), as GP-included BHA-PI does in Figure 8(d)-(f). The corresponding MES-PIMD results of the same model are shown in Figure 4.

Fitting of the results produced for sufficiently large values of $N_{\text{bead}}$ reveals that the numerical convergence of the Trotter decomposition is at $O\left(1/\sqrt{N_{\text{bead}}}\right)$ for BHA-PI, GP-included BO-PI/BHA-PI constructed by brute-force incorporation of the winding-number-induced phase factor, SES-overlap-PI, and GP-excluded SES-overlap-PI obtained through brute-force incorporation of the winding-number-induced phase factor. In contrast, BO-PI yields conventional $O\left(1/N_{\text{bead}}^2\right)$ numerical convergence of the Trotter decomposition. (See Table S2 and SI Section S2.7.3) Similar to the previous discussion on the GP-*excluded* Hamiltonian (eq (4)) and GP-*excluded* MES-PI results, *incorporation* of the GP into the BO Hamiltonian (i.e., eq (22), GP-included BO Hamiltonian) leads to a singular behavior at the origin ( $r=0$ ), as illustrated by the cusp-like structure, the divergent derivative and the strictly vanishing probability at the origin of the ground-state wavefunction (Figure S4(a, b, and d)). This singularity destroys the second-order error cancellation of the Trotter decomposition for PI, where the winding-number-induced phase factor is incorporated by brute-force. For the BHA Hamiltonians, the DBOC term, $1/8r^2$, also introduces a "cusp" or singularity at the origin (Figure S5). The anomalous convergence rate can be rationalized by invoking the state-dependent bounds associated with the Trotter factorization (see ref [65] for the discussion in a different context). Figure 8 and Figure 9, as well as Table S2 of the Supporting Information, demonstrate that the GPA-SP algorithm improves SES approaches constructed by brute-force incorporation of the winding-number-induced phase factor, resulting in considerably faster convergence.

Throughout this Perspective, we have used PIMD (with the efficient "middle" scheme [24, 66-72]) for all MES-PI and SES-PI simulations because PIMD is applicable to general complex molecular systems. For instance, it is straightforward to extend our recently developed efficient PIMD method for the isobaric-isothermal ensemble [68] into the MES-PI, SES-overlap-PI, and GP-included BHA-PI/BO-PI approaches to study realistic nonadiabatic systems under conditions of controlled pressure and temperature. We also note that PI Monte Carlo[48] offers a viable alternative when it involves only modest work to determine the Metropolis step sizes for all nuclear DOFs.

For demonstration purposes throughout this Perspective, we employ the linear $E \otimes e$ Jahn-Teller model, whose adiabatic potential-energy surfaces are cylindrically symmetric. We highlight a key physical regime: when the thermal de Broglie wavelength becomes comparable to the geometric length scale required for the closed imaginary-time path to encircle the CI, the GP profoundly alters thermodynamic properties. This fundamental criterion is not limited to the linear Jahn-Teller model; it remains valid for more complex systems, including Jahn-Teller Hamiltonians with quadratic coupling terms and general coupled MES systems involving multi-nuclear DOFs and multi-CI seams.

**General Systems in the Adiabatic Representation.** The adiabatic representations for the two-state model Hamiltonians in eq (19) are obtained from the diabatic presentation of eq (1), which implies a complete Hilbert space. In such a space, a pure gauge is available, namely, the gauge field tensor (as described by eq (S34) of the Supporting Information[73] of ref [74], eq (12) of ref [40], and in refs [7, 75]),

$$\hat{\mathcal{F}}^{(KL)} = \frac{\partial \mathbf{A}^{(L)}}{\partial R_K} - \frac{\partial \mathbf{A}^{(K)}}{\partial R_L} + \mathrm{i}\left[\mathbf{A}^{(K)}, \mathbf{A}^{(L)}\right], \tag{31}$$

vanishes identically [76-78] for the Yang-Mills field/non-Abelian gauge field, $\mathbf{A}^{(K)} = \mathbf{A}_{\mathrm{GP}}^{(K)} - \mathrm{i}\mathbf{d}^{(K)}$. Here, matrix $\mathbf{A}^{(K)}$ is the $K$-th nuclear component of the tensor, $\mathbf{A} = \mathbf{A}_{\mathrm{GP}} - \mathrm{i}\mathbf{d}$. In addition, the nonadiabatic scalar coupling matrix, defined as $D_{mn}^{(K)}(\mathbf{R}) = \left\langle \phi_m(\mathbf{R}) \left| \frac{\partial^2}{\partial R_K^2} \right| \phi_n(\mathbf{R}) \right\rangle$, is related to $\mathbf{A}^{(K)}$ through

$$\delta \mathbf{D}^{(K)} = \mathbf{D}^{(K)} + \mathbf{A}^{(K)}\mathbf{A}^{(K)} - \mathrm{i}\frac{\partial}{\partial R_K}\mathbf{A}^{(K)} \quad , \tag{32}$$

which also vanishes identically in a complete Hilbert space [40, 77, 78]. (Equation (32) is equivalent to eq (11) of ref [40].)

However, practical simulations of real molecular systems necessarily involve a truncation of the electronic Hilbert space, retaining only the lowest $F$ adiabatic electronic states out of an infinite manifold. When $\mathbf{A}^{(K)}$ is defined within this truncated subspace, eqs (31) and (32) generally acquire nonzero matrix elements [40, 78], given by,

$$\mathcal{F}^{(KL)}_{mn} = -\mathrm{i}\sum_{r=F+1}^{+\infty}\left(A^{(K)}_{mr}A^{(L)}_{rn} - A^{(L)}_{mr}A^{(K)}_{rn}\right), \quad \delta D^{(K)}_{mn} = -\sum_{r=F+1}^{+\infty} A^{(K)}_{mr}A^{(K)}_{rn}, \tag{33}$$

which reflect the incompleteness of the truncated subspace and its residual coupling to the discarded higher-lying electronic-states. For a reasonable truncation, both $\delta\mathbf{D}^{(K)}$ and $\mathcal{F}^{(KL)}$ are expected to remain small. Notably, $\delta\mathbf{D}^{(K)}$ contains negative-definite quadratic forms $\delta D^{(K)}_{nn} = -\sum_{r=F+1}^{\infty} |A^{(K)}_{rn}|^2$, which constrain the magnitudes of the couplings to the higher-lying electronic states and, consequently, bound the gauge field tensor. Thus, $\delta\mathbf{D}^{(K)} = 0$ (for all $K$) constitutes a sufficient condition for $\mathcal{F}^{(KL)} = 0$, whereas the converse does not generally hold. SI Section S3 demonstrates that both $\delta\mathbf{D}^{(K)}$ and $\mathcal{F}^{(KL)}$ can be extracted from products of the truncated $F \times F$ overlap matrices along an infinitesimal square loop (similar to the Wilson loop in a different context in ref [79]).

We further demonstrate that, under appropriate conditions, the error arising from the truncation of the electronic basis set is systematically controllable. Consider the low-temperature regime defined by $k_{\mathrm{B}}T \ll \Delta E$, where $\Delta E$ is the representative energy gap between the potential minimum of the lowest discarded electronic-state and that of the highest retained electronic-state. The error introduced by performing MES-PI with truncated overlap matrices and potential-energy matrices can be estimated by expanding the electronic trace of the weighted overlap product along the ring polymer. The leading contributions to the truncation error arise from terms involving a virtual transition to the lowest discarded state from the retained manifold. It is straightforward to show the overall dominate component is $\mathrm{e}^{-\beta\Delta E}$. Therefore, provided the thermal energy $k_{\mathrm{B}}T$ is low relative to the energy gap $\Delta E$, MES-PI simulations confined to the truncated electronic subspace yield a reliable and convergent description of thermodynamic properties. For instance, in the Jahn-Teller model considered in this Perspective, when $\Delta E$, chosen as $E_{\mathrm{JT}} = c^2/2$ (defined in Figure 1(a)), is significantly larger than the thermal energy $k_{\mathrm{B}}T = \beta^{-1}$, the error induced by truncating the electronic Hilbert space from two states to a single ground state is negligible. This behavior is illustrated in Figure 6(c-d) with approximately $k_{\mathrm{B}}T < 1$ r.u., i.e., $E_{\mathrm{JT}}/k_{\mathrm{B}}T > 4.5$, where the solid red line (the basis-set-expansion results of the MES Hamiltonian, which are equivalent to the MES-PI results, in the full electronic space) closely coincides with the dotted red line (representing results from the corresponding calculations in the truncated electronic subspace). This agreement confirms that the GP effect in thermodynamic properties is accurately captured by MES-PI even when higher-energy electronic states are omitted in the regime $k_{\mathrm{B}}T \ll \Delta E$.

**Concluding Remarks.** In summary, we demonstrate that the GP effect significantly influences thermodynamic properties, particularly in the low-temperature regime. We have shown that the MES-PI approach[24] (i.e., eq (14)) offers a rigorous and natural framework for incorporating the GP in simulations of thermodynamic properties. In MES-PI, the GP arises intrinsically from the electronic trace of the product of (statistically weighted) overlap matrices along the ring polymer. We construct an *ad hoc* GP-excluded MES-PI approach to unambiguously isolate purely topological effects from other nonadiabatic contributions. This is achieved by introducing a geometric signature matrix, which accounts for both topologically non-trivial and trivial CIs, in conjunction with a winding-number-induced phase factor. In the SES limit where excited electronic states are well separated in energy and the temperature is sufficiently low, the SES-overlap-PI scheme, i.e., the SES limit of MES-PI, offers a simpler practical tool for including the GP effect, albeit at the cost of slower numerical convergence (e.g., a reduced Trotter convergence order). Notably, both MES-PI and SES-overlap-PI (illustrated in Figures 2(a) and 7(c), respectively) remain applicable to complex/large systems where the location and topology of CIs are not known *a priori*. When the SES limit is valid and the CIs are known, by incorporating a winding-number-induced phase factor, GP-included BHA-PI or GP-included BO-PI (illustrated in Figure 7(b)) correctly accounts for the GP in low-temperature thermodynamics. Moreover, the GPA-SP algorithm proposed in SI Section S2.5.2 addresses the associated convergence challenges, substantially improving the efficiency of GP-included BHA-PI/BO-PI approaches. Although demonstrated on a prototype model, the results presented in this Perspective establish a framework for future investigations into thermodynamic properties of real nonadiabatic molecular systems, such as single-molecule magnets and ultracold molecules. Furthermore, by providing a rigorous description of nonadiabatic systems at thermal equilibrium, the MES-PI framework establishes the initial conditions necessary for real-time simulations of such processes. Integrating MES-PI with the nonadiabatic field approach[40, 80, 81] formulated in the generalized coordinate–momentum phase space of quantum mechanics[74, 82-88], symmetrized quasi-classical Meyer-Miller mapping Hamiltonian approach[89-94], surface hopping[95-99], *ab initio* multiple spawning[100, 101], exact factorization[102, 103], small matrix path integral[104-106], and other nonadiabatic dynamics methods[107, 108], will facilitate the study of complex thermally activated or light-driven nonadiabatic dynamics in real molecular systems. Lastly, the strategies of the imaginary-time PI formulation can be straightforwardly extended to the real-time PI formulation for MES systems. For example, The correspondence between real and imaginary time slices is established *via* the Wick rotation $\beta = \mathrm{i}t/\hbar$. For a total duration *t* discretized into $N_{\mathrm{step}}$ intervals, each time slice is given by $t/N_{\mathrm{step}}$. The product of "weighted" (electronic) overlap matrices between successive time slices, $\prod_{j=1}^{N_{\mathrm{bead}}} \mathrm{e}^{-\beta \mathbf{\Lambda}^{[j]}/N_{\mathrm{bead}}} \mathbf{C}^{[j,j+1]}$ in imaginary-time or $\prod_{j=1}^{N_{\mathrm{step}}} \mathrm{e}^{-\mathrm{i}t\mathbf{\Lambda}^{[j]}/\hbar N_{\mathrm{step}}} \mathbf{C}^{[j,j+1]}$ in real-time, is the key to capture the GP effect in real complex systems where the location and topology of CI seams are often not known *a priori*. This critical point **was already implicit** in the benchmark MES-PIMD simulations of the modified seven-state FMO model reported in ref [24] (see the model defined by eq (80) in ref [24] ), which did not enforce sign continuity of the electronic eigenvectors (i.e., the eigenvectors from the diagonalization of DPEM) along successive imaginary-time slices. We anticipate that this Perspective, along with the work of ref [24], will provide a foundation of new tools for addressing the increasingly important role of topological effects in quantum statistical mechanics and nonadiabatic dynamics for real systems.[6-8, 109]

ASSOCIATED CONTENT

**Supporting Information**.

Supporting Information is available free of charge via the Internet at the ACS website.

Bound states of the Jahn-Teller $E\otimes e$ model; path integral simulation; gauge field tensor and nonadiabatic scalar-coupling corrections from truncation (PDF)

Comparison of the heat capacity for the Jahn–Teller $E\otimes e$ model calculated using different theoretical approaches as the parameter, $c$, varies (GIF)

**AUTHOR INFORMATION**

**Corresponding Author**

*E-mail: jianliupku@pku.edu.cn

**ORCID**

Yu Zhai: 0000-0002-5065-688X

Youhao Shang: 0000-0002-9297-6654

Jian Liu: 0000-0002-2906-5858

**Notes**

The authors declare no competing financial interests.

## ACKNOWLEDGMENT

This work was supported by the National Science Fund for Distinguished Young Scholars Grant No. 22225304, and the open research fund of Key Laboratory of Precision and Intelligent Chemistry (KY2490000300). We acknowledge the High-performance Computing Platform of Peking University, Beijing PARATERA Tech Co., Ltd., and Guangzhou supercomputer center for providing computational resources. We also thank the Laoshan Laboratory (LSKJ202300305) for providing the computational resources of the new Sunway platform and for the technical support.

## ABBREVIATIONS

| **Abbreviations** | **Complete Terminology** |
|---|---|
| BHA | Born-Huang Adiabatic |
| BO | Born-Oppenheimer |
| CI | conical intersection |
| DBOC | diagonal Born-Oppenheimer correction |
| DOF | degree of freedom |
| DPEM | diabatic potential energy matrix |
| GP | geometric phase |
| GPA-SP | GP absorbed into the spring potential |
| MES | multi-electronic-state |
| PES | potential energy surface |
| PI | path integral |
| PIMD | path integral molecular dynamics |
| r.u. | reduced unit |
| SES | single-electronic-state |